\documentclass[11pt,letterpaper]{article}

\usepackage{amsmath,amsfonts,amssymb,amsthm}
\usepackage{deepthink}
\usepackage{mathrsfs}
\usepackage{mathtools}
\usepackage{multirow}
\usepackage{manyfoot}
\usepackage{booktabs}
\usepackage{algorithm,algpseudocode}
\usepackage{array}
\usepackage{textcomp}
\usepackage{stfloats}
\usepackage{xurl}
\usepackage{verbatim}
\usepackage{colortbl}
\usepackage{ifthen}

\usepackage{url}

\hypersetup{
    colorlinks=true,
    linkcolor=blue,
    citecolor=blue,      
    urlcolor=blue,
    pdfpagemode=FullScreen,
}

\usepackage{amsmath,amsthm,amssymb,amsbsy,amsfonts,amscd,bm}
\usepackage{xcolor}
\usepackage{color}
\usepackage{graphicx}
\graphicspath{{./figs/}}
\usepackage{comment}
\usepackage{multirow}
\usepackage{enumitem}
\usepackage{fancyhdr}
\usepackage{subcaption}
\usepackage{wrapfig}
\usepackage{needspace}

\newcommand{\e}{\begin{equation}}
\newcommand{\ee}{\end{equation}}
\newcommand{\en}{\begin{equation*}}
\newcommand{\een}{\end{equation*}}
\newcommand{\eqn}{\begin{eqnarray}}
\newcommand{\eeqn}{\end{eqnarray}}
\newcommand{\bmat}{\begin{bmatrix}}
\newcommand{\emat}{\end{bmatrix}}

\DeclareMathAlphabet\mathbfcal{OMS}{cmsy}{b}{n}

\newcommand{\mb}{\bm}
\newcommand{\mc}{\mathcal}

\DeclareMathOperator*{\argmin}{\text{arg~min}}

\newcommand{\wh}{\widehat}

\graphicspath{{./figs/}}

\newlength{\imgwidth}
\newboolean{twoColVersion}
\setboolean{twoColVersion}{false}
\newcommand{\twoCol}[2]{\ifthenelse{\boolean{twoColVersion}} {#1} {#2} }

\newcommand{\qq}[1]{\textcolor{blue}{\bf [{\em Qing:} #1]}}

\usepackage[
  backend=biber,
  style=alphabetic,
  maxbibnames=100,
  minbibnames=100,
  maxcitenames=2,
  mincitenames=2
]{biblatex}
\usepackage[nameinlink]{cleveref}

\definecolor{LightCyan}{rgb}{0.88,1,1}
\definecolor{Columnbia Blue}{rgb}{0.607,0.866,1}
\graphicspath{{figures/}}

\title{DCDP: Decoupled Data Consistency via Diffusion Purification for Solving General Inverse Problems}

\authorblock{
  \textbf{Xiang Li}\textsuperscript{1},
  \textbf{Soo Min Kwon}\textsuperscript{1},
  \textbf{Shijun Liang}\textsuperscript{2},
  \textbf{Ismail Alkhouri}\textsuperscript{1},
  \textbf{Saiprasad Ravishankar}\textsuperscript{2},
  \textbf{Qing Qu}\textsuperscript{1}
}

\affiliation{
  \textsuperscript{1}Department of Electrical Engineering and Computer Science,
  University of Michigan\par
  \textsuperscript{2}Departments of Computational Mathematics, Science and Engineering,
  and Biomedical Engineering, Michigan State University
}

\authornote{
  \textsuperscript{1}\texttt{\{forkobe,kwonsm,ismailal\}@umich.edu}
  \quad
  \textsuperscript{2}\texttt{\{liangs16,ravisha3\}@msu.edu}
}

\abstracttext{
Diffusion models have emerged as powerful generative priors, excelling in solving inverse problems owing to their ability to model complex data distributions. Existing methods typically enforce data consistency by guiding the reverse sampling process with additional likelihood gradients. However, computing these gradients introduces significant computational overhead, prolongs inference time, and most importantly constrains compatibility with accelerated samplers. To overcome these limitations, we propose DCDP (Decoupled Data Consistency via Diffusion Purification), a two-stage diffusion-based framework for solving inverse problems that disentangles data consistency from reverse sampling steps. Specifically, DCDP iteratively alternates between the following two phases: (\emph{i}) a reconstruction phase that enforces data consistency, followed by (\emph{ii}) a refinement phase that applies diffusion purification to impose the learned generative prior. DCDP is broadly applicable to pixel space and latent space diffusion, and consistency models, achieving significant speedup in inference time compared to most state-of-the-art methods. We demonstrate its effectiveness and generality across a range of inverse problems, including image denoising, (nonlinear) deblurring, inpainting, and super-resolution.
}

\date{Originally released on arXiv in March 2024; revised in October 2026}
\correspondence{\href{mailto:forkobe@umich.edu}{forkobe@umich.edu}}
\resources{\href{https://github.com/Morefre/Decoupled-Data-Consistency-with-Diffusion-Purification-for-Image-Restoration}{Code}}
\headerlogo{}{}

\begin{document}

\makeDeepthinkHeader
\begin{figure*}[h]
    \centering
    \includegraphics[width=0.9\textwidth]{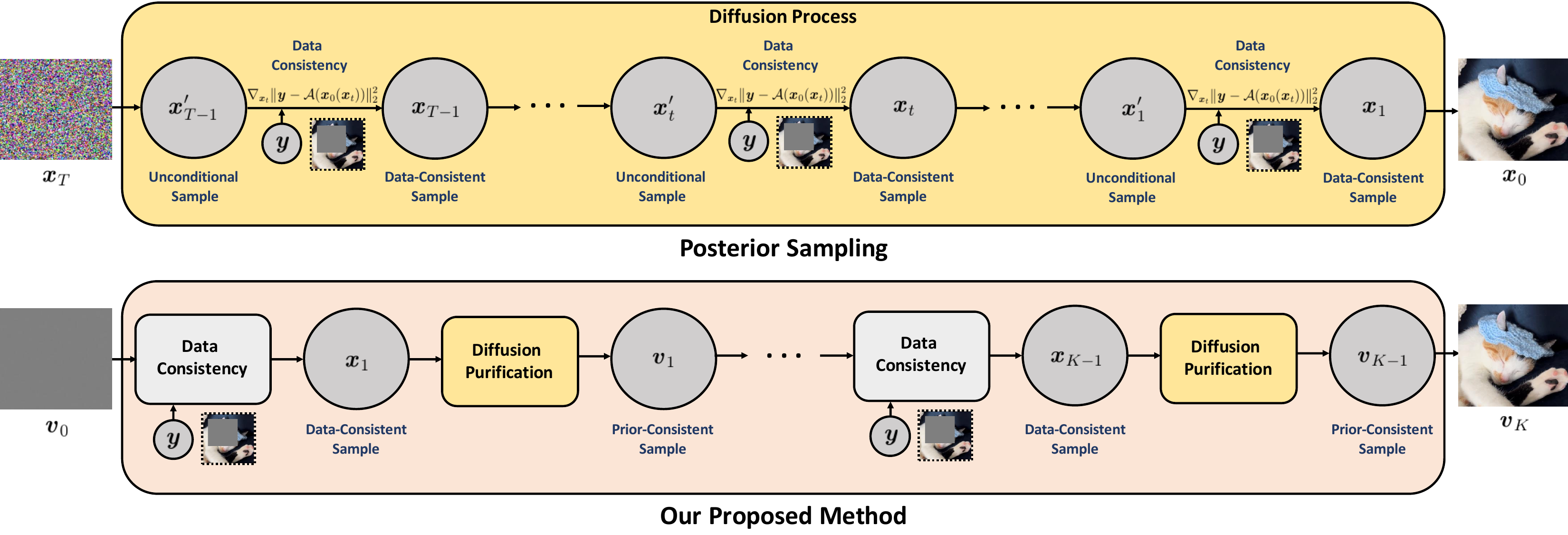}
    \caption{\textbf{Illustration of the difference between our proposed method and existing posterior sampling techniques for solving inverse problems.} Our method decouples data consistency from the diffusion process, unlike existing approaches that enforce it via gradient steps during reverse sampling. }
    \label{fig:teaser}
\end{figure*}
\tableofcontents
\newpage

\section{Introduction}
\label{sec: intro}

Inverse problems naturally arise in many image processing and scientific applications, such as image denoising, deblurring, inpainting, super-resolution, and phase retrieval~\cite{ongie2020deep, ravishankar2019image, zbontar2018fastmri}. Mathematically, the goal of solving an inverse problem is to estimate an unknown signal $\mb x^\star \in \mathbb R^n$ from its subsampled (and possibly corrupted) measurements $\mb y\in \mathbb R^m$:
\begin{align}
    \label{eq: forward measurement process}
    \mb y \;=\; \mathcal{A}(\mb x^\star) + \mb n ,
\end{align}
where $\mathcal{A}: \mathbb{R}^n \rightarrow \mathbb{R}^m$ is a linear or nonlinear forward operator and $\mb n \in \mathbb R^m$ is the additive noise. Typically, these inverse problems are highly ill-posed with $m \ll n$, resulting in infinitely many solutions that satisfy \eqref{eq: forward measurement process}. 

A common approach to address this challenge is to regularize the solution by enforcing certain data priors. In the past, many classical handcrafted priors such as sparsity or total variation have been investigated in the literature~\cite{ravishankar2019image}. 
However, handcrafted priors are often overly simplistic, limiting their ability to capture the complex structures inherent in natural images.
To deal with the challenge, deep generative models have emerged as better priors, significantly outperforming handcrafted priors in solving inverse problems. This paradigm shift was pioneered by recent seminal work such as~\cite{bora2017compressed}, which showed that deep generative priors enable superior image reconstruction with fewer measurements than traditional sparsity-based methods; see also subsequent developments~\cite{ongie2020deep}.

More recently, diffusion models have gained increasing attention as image priors~\cite{ho2020denoising, song2020score, dhariwal2021diffusion, song2020denoising}. These models represent a novel class of deep generative models that are capable of producing high-quality samples thanks to their powerful ability to model complex data distributions. Specifically, diffusion models operate by learning to reverse a forward process that progressively adds Gaussian noise to the data till it becomes indistinguishable from pure noise. The key to performing the reverse process is learning the score functions $\nabla_{\bm{x}_t}\log p(\bm{x}_t)$, i.e., the gradients of the log-density of the noise-mollified data $\bm{x}_t$, which can be approximated with deep networks trained with the denoising score matching objective~\cite{vincent2011connection}.

When applied to solving inverse problems, the reverse process needs to be guided by additional likelihood gradients $\nabla_{\bm{x}_t}\log p(\bm{y}|\bm{x}_t)$ to ensure consistency with the measurements $\mb y$.
However, since the likelihood gradient $\nabla_{\bm{x}_t}\log p(\bm{y}|\bm{x}_t)$ is generally intractable, existing methods either iteratively project $\bm{x}_t$ onto the measurement subspace~\cite{song2020score, chung2022come, chung2022improving, choi2021ilvr, wang2022zero}, or approximate the gradient under simplified assumptions~\cite{chung2022diffusion, song2023PiGDM, jalal2021robust}. These approaches face fundamental challenges due to the coupling of data consistency and the reverse sampling process. Specifically, data consistency is typically enforced at each diffusion step, so that the total number of consistency updates is constrained by the number of reverse steps. As a result, achieving high reconstruction quality often requires a large number of diffusion steps, limiting compatibility with accelerated samplers such as Denoising Diffusion Implicit Models (DDIMs)~\cite{song2020denoising} or consistency models (CMs)~\cite{song2023consistency}, which operate with significantly fewer steps by improved designs. Moreover, this coupling complicates the extension to latent diffusion models: recent works~\cite{rout2024solving, song2023solving} demonstrate that naively incorporating the likelihood gradient fails to ensure data consistency in the latent space, posing a significant challenge for fast and accurate inference in high-dimensional inverse problems \cite{chungdecomposed}.

\medskip
\noindent \textbf{Our Contributions.} In this study, we tackle the challenges of solving general inverse problems with diffusion models by introducing a new method termed Decoupled Data Consistency via Diffusion Purification (DCDP), which decouples data consistency steps from the reverse sampling process. As illustrated in \Cref{fig:teaser}, our algorithm iteratively alternates between: (\emph{i}) a reconstruction phase that enforces measurement consistency by minimizing a data fidelity objective, and (\emph{ii}) a refinement phase that refines the estimate using a diffusion-based image prior via \emph{diffusion purification}~\cite{nie2022diffusion}. Importantly, the data consistency update is executed independently of the reverse diffusion process, offering two key advantages over existing diffusion-based inverse problem solvers: 
\begin{itemize}[leftmargin=*]
    \item \textbf{Versatility.} Our framework can seamlessly integrate a broad range of diffusion models in a plug-and-play fashion, including latent diffusion for improved memory efficiency and consistency models for accelerated sampling.
    \item \textbf{Efficiency.} 
    By decoupling these processes, we can significantly reduce the inference time (see \Cref{fig:enter-label}), while achieving performance comparable to state-of-the-art algorithms.
\end{itemize}
Empirically, we validate our method on both linear (super-resolution, inpainting, and image deblurring) and nonlinear inverse problems (nonlinear deblurring) across various datasets including FFHQ~\cite{karras2019style}, LSUN-Bedroom~\cite{yu2015lsun}, and ImageNet~\cite{deng2009imagenet},  achieving competitive results over existing methods with a faster inference speed.


\medskip
\noindent \textbf{Notations.} Throughout this paper, we denote scalars (functions) with regular lower-case letters (e.g., $s_t, f(t)$) and vectors (vector functions) with bold lowercase letters (e.g., $\bm x, \bm s(\bm x_t, t)$). We use $[N]$ to denote the set $\{1, 2, \ldots, N\}$,  $p(\cdot)$ to denote the probability, $\mathbb E\left[\cdot\right]$ to denote the expectation, $\|\cdot\|_2$ to denote the $\ell_2$ norm, and we use $\mathcal N(\cdot,\cdot)$ to denote the Gaussian distribution. We use $\bm{x}_t$ to denote a sample $\bm{x}$ at time step $t$. We use $\mb x_0$ to denote a clean sample, and $\mb s_{\mb{\theta}}$ and $\mb s_{\mb{\phi}}$ for the estimated score functions in the pixel and latent spaces, respectively. 

 

\section{Background}
\label{sec: background}

\subsection{Unconditional Diffusion Models}
\label{subsec:background-dm}
Diffusion models~\cite{sohl2015deep, ho2020denoising, song2020score, dhariwal2021diffusion} generate samples by reversing a pre-defined forward process. The forward process iteratively diffuses the data distribution $p_0(\mb x)$ with Gaussian noise until a sufficiently large timestep $T$, so that the noise-mollified data distribution $p_T(\mb x)$ becomes indistinguishable from an isotropic Gaussian distribution with covariance $\sigma^2_T\mb I$. In the variance-preserving (VP) setting \cite{song2020score}, the forward process can be written as a stochastic differential equation (SDE) of the form
\begin{align}
    \label{eq: forward SDE}
    d\mb x_t \;=\; -\frac{\beta_t}{2}\mb x_t dt + \sqrt{\beta_t}d\mb w,
\end{align}
where $\beta_t>0$ controls the speed of diffusion and $\mb w$ follows the standard Wiener process. To sample an image, we can start from $\mb x_T\sim \mathcal{N}(\mb 0,\sigma^2_T\mb I)$ and reverse the forward process~\eqref{eq: forward SDE}. As shown in~\cite{song2020score}, the forward process~\eqref{eq: forward SDE} can be reversed with either a reverse SDE:
\begin{align} 
    d\mb x_t \;=\;& \left [-\frac{\beta_t}{2}{\mb{x}_t} -\beta_t\nabla_{\mb x_t}\log p_t(\mb x_t) \right]dt 
    + \sqrt{\beta_t}d\bar{\mb w}, \label{eq: Reverse SDE}
\end{align}
or a reverse ordinary differential equation (ODE):
\begin{align} 
    d\mb x_t \;=\;& \left [-\frac{\beta_t}{2}{\mb{x}_t} -\frac{\beta_t}{2}\nabla_{\mb x_t}\log p_t(\mb x_t) \right]dt , \label{eq: Reverse ODE}
\end{align}
where $p_t(\bm x_t)$ is the intermediate noise-mollified distribution at time $t$, $\nabla_{\mb x_t}\log p_t(\bm x_t)$ is the score function, and $\bar{\mb w}$ is the standard Wiener process~\cite{anderson1982reverse}. In practice, the score function is estimated using a deep neural network $\mb s_{\bm\theta}$ with parameter $\mb{\theta}$ trained using the weighted denoising score matching objective~\cite{vincent2011connection}:
\begin{equation}
  \scalebox{1}{$ 
    \underset{\mb\theta}{\min}\;
    \mathbb{E}_{t,\mb x_0,\mb x_t}
    \bigl[\lambda(t)\|\mb s_{\mb\theta}(\mb x_t,t)
      -\nabla_{\mb x_t}\log p_{0t}(\mb x_t\mid \mb x_0)\|_2^{2}\bigr],
  $}
\end{equation}
where $\lambda(t)$ is a positive weighting function, $t\sim \mathcal{U}(0,1)$, $\mb x_0\sim p_0(\mb x)$, and $\mb x_t\sim p_{0t}(\mb x_t| \mb x_0)$, where $p_{0t}(\mb x_t|\mb x_0)$ is the transition kernel, i.e., the distribution of $\mb x_t$ induced by~\eqref{eq: forward SDE} given $\mb x_0$. In the VP setting, the transition kernel takes the following form: 
\begin{align}
     p_{0t}(\mb x_t|\mb x_0)=\mathcal{N}(\sqrt{\alpha_t}\mb x_0,(1-\alpha_t)\mb I),
\end{align}
where $\alpha_t=e^{-\int_{0}^t\beta_sds}$.
\subsection{Diffusion Posterior Sampling for Solving Inverse Problems}
To solve inverse problems using diffusion models, one of the most commonly used techniques is to modify the unconditional reverse sampling process in \eqref{eq: Reverse SDE} by replacing the unconditional score with a conditional score function $\nabla_{\mb x_t}\log p_t(\mb x_t|\mb y)$ conditioned on measurement $\mb y$. As such, the conditional reverse SDE has the following form
\begin{align}
\label{eq: Conditional Reverse SDE}
 d\mb x_t \;=\; \left [-\frac{\beta_t}{2}{\mb{x}_t} -\beta_t\nabla_{\mb x_t}\log p_t(\mb x_t|\mb y) \right]dt +\sqrt{\beta_t}d\bar{\mb w}.
\end{align}
Using Bayes rule, the conditional score function $\nabla_{\mb x_t}\log p_t(\mb x_t|\mb y)$ can be decomposed into the unconditional score and the likelihood gradient: 
\begin{align}
    \nabla_{\mb x_t}\log p_t(\mb x_t|\mb y)=\underbrace{\nabla_{\mb x_t}\log p_t(\mb x_t)}_{\text{unconditional score}} + 
    \underbrace{\nabla_{\mb x_t}\log p_t(\mb y|\mb x_t)}_{\text{likelihood gradient}}.
\end{align}
While the unconditional score $\nabla_{\mb x_t}\log p_t(\mb x_t)$ can be approximated using a pre-trained diffusion model $\mb s_{\mb\theta}(\mb x_t,t)$ (see \Cref{subsec:background-dm}), the likelihood gradient $\nabla_{\mb x_t}\log p_t(\mb y|\mb x_t)$ is computationally intractable in general. In the past, many existing methods focused on estimating the likelihood  gradient under simplified assumptions~\cite{chung2022diffusion, song2023PiGDM, jalal2021robust}. Among them, one of the most popular approaches is diffusion posterior sampling (DPS)~\cite{chung2022diffusion}, which approximates the likelihood gradient as 
\begin{align}
\label{eq: DPS likelihood approximation}
    \nabla_{\mb x_t}\log p_t(\mb y|\mb x_t) \approx \nabla_{\mb x_t}\|\mathcal{A}(\hat{\mb x}_0(\mb x_t))-\mb y\|_2^2,
\end{align}
where $\hat{\mb x}_0 (\mb x_t) := \mathbb{E}[\mb x_0|\mb x_t]$ is the posterior mean, which can be computed with the unconditional score function through Tweedie's formula:
\begin{align}
    \hat{\mb x}_0(\mb x_t) 
    &= \frac{1}{\sqrt{\alpha_t}}(\mb x_t+(1-\alpha_t)\nabla_{\mb x_t}\log p(\mb x_t))\\
    &\approx \frac{1}{\sqrt{\alpha_t}}(\mb x_t+(1-\alpha_t)\mb s_{\mb \theta}(\mb x_t,t)).
\end{align}
Note the $\ell_2$ norm in~\eqref{eq: DPS likelihood approximation} is a direct consequence of assuming Gaussian measurement noise; if the noise follows a different distribution, the norm can be replaced accordingly. Although effective, the DPS-based method still suffers from several limitations: (\emph{i}) the number of data consistency steps is limited by the number of reverse sampling steps, potentially resulting in insufficient measurement consistency when employing accelerated samplers, and (\emph{ii}) computing the approximated likelihood gradient~\eqref{eq: DPS likelihood approximation} requires backpropagating through the entire score function $\mb s_{\mb\theta}(\mb x_t,t)$ at each timestep $t$, which is computationally expensive since $\mb s_{\mb\theta}(\mb x_t,t)$ is typically parameterized by a large deep network. Moreover, adapting DPS to latent diffusion models (LDMs)~\cite{rombach2022high} is highly nontrivial due to the nonlinear nature of their encoders and decoders \cite{song2023solving,rout2024solving}.

\subsection{Diffusion Purification}
\label{sec:dpur}
Recently, diffusion purification has been shown to be an effective approach for removing adversarial perturbations~\cite{nie2022diffusion, alkhouri2023robust}. Specifically, for a given adversarially perturbed image $\mb x_\textrm{adv}$, diffusion purification first runs the forward process~\eqref{eq: forward SDE} starting from $\mb x_\textrm{adv}$ to some intermediate timestep $t^{\prime}$ such that $\mb x_{t^{\prime}}\sim p_{0t^{\prime}}(\mb x_{t^{\prime}}|\mb x_\textrm{adv})$: 
\begin{align}\label{eqn:diff-pur-adv}
    \mb x_{t^{\prime}} = \sqrt{\alpha_{t^{\prime}}}\mb x_\textrm{adv}+\sqrt{(1-\alpha_{t^{\prime}})}\mb \epsilon\:,
\end{align}
where $\mb\epsilon \sim\mathcal{N}(\mb 0,\mb I)$. Starting from $\mb x_{t^{\prime}}$, diffusion purification performs the reverse diffusion process~\eqref{eq: Reverse SDE} or~\eqref{eq: Reverse ODE} to produce a ``purified'' image $\mb x_\textrm{purified}$, effectively eliminating the adversarial perturbations. The rationale behind this approach is that by carefully selecting $t^{\prime}$, the adversarial noise within $\mb x_\textrm{adv}$ becomes overwhelmed by the noise introduced during the forward process, yet the core semantic and structural information is preserved. Consequently, executing the reverse diffusion process leads to a ``clean'' image reconstruction.

Distinct from existing methods that use diffusion purification to improve the robustness of DNN-based classifiers~\cite{nie2022diffusion}, in our work, we use diffusion purification as an operator that enforces the image prior embedded in the diffusion models.

\section{Proposed Method}\label{sec: method}
This section first motivates DCDP from the perspective of classical variable splitting techniques (\Cref{sec:motivation}). Then we present our overall algorithm (\Cref{sec:dcdp}) and its extension to latent diffusion models and consistency models (\Cref{subsec:fast-sampling,sec: adaptation to LDMs}). 
\subsection{Motivation: Methods with Variable Splitting}
\label{sec:motivation}

Our method is inspired by the classical variable splitting techniques \cite{afonso2010fast, boyd2011distributed, venkatakrishnan2013plug, zhang2021plug} in optimization. Consider the following regularized optimization formulation for solving~\eqref{eq: forward measurement process},
\begin{align*}
    \min_{\mb x} f(\mb x) + \lambda \mathcal R(\mb x), 
\end{align*}
where we typically adopt the least squares formulation, $f(\mb x) := \frac{1}{2} \| \mathcal A(\mb x) - \mb y \|_{2}^2$, as the data fidelity term to enforce consistency with the measurements, $\mc R(\cdot)$ is the regularizer that enforces the data prior, and $\lambda>0$ is a hyperparameter that balances the fidelity and the regularization strength. 
The idea behind variable splitting is to introduce an auxiliary variable $\mb v = \mb x$, resulting in an equivalent problem
\begin{align}\label{eq: constrained}
    \min_{\mb x, \mb v} f(\mb x) + \lambda \mc R(\mb v), \quad  \mathrm{s.t.}~~\; \mb x = \mb v,
\end{align}
where the constrained optimization problem above can be further relaxed into an unconstrained problem by the half quadratic splitting technique (HQS)~\cite{geman1995nonlinear, zhang2021plug}:
\begin{equation}\label{eqn:augmented}
\min_{\mb x,\mb v}
\mathcal L(\mb x,\mb v):=f(\mb x)+\lambda\,\mathcal R(\mb v)+\mu\|\mb x-\mb v\|_2^{2},
\end{equation}
for some properly chosen $\mu>0$. To solve~\eqref{eqn:augmented}, a common strategy is to adopt an iterative method that, at each iteration $k$, uses the auxiliary variable $\mb v$ to decompose the optimization into two subproblems of the form:
\begin{align}
    \label{hqs_data_fidelity}
        &\mb x_k = \arg\min_{\mb x} \; \mc L(\mb x, \mb v_{k-1}) , \\
    \label{hqs_diffusion}
        &\mb v_k = \arg\min_{\mb v} \; \mc L(\mb x_k, \mb v).
\end{align}
In this way, the data-consistency term $f(\mb x)$ is \emph{decoupled} from the regularization term $\mc R(\mb v)$. Then, instead of having an explicit regularizer $\mathcal{R(\cdot)}$ (e.g., $\ell_1$-norm to induce sparsity), the regularization-based step~\eqref{hqs_diffusion} can be replaced with operators that \emph{implicitly} enforce the data prior. In this work, we leverage diffusion purification to enforce the image prior encoded within the diffusion models as we discuss in \Cref{sec:dcdp}.

\subsection{Decoupled Data Consistency via Diffusion Purification}
\label{sec:dcdp}
Our DCDP method alternates between the following two phases: (\emph{i}) a reconstruction phase that enforces data consistency and (\emph{ii}) a refinement phase that enforces the prior via diffusion purification. Below, we discuss these steps in more detail, with an illustration in Figure~\ref{fig:teaser}.

\medskip
\noindent \textbf{Reconstruction Phase.}
The first step of our algorithm is to solve the sub-problem~\eqref{hqs_data_fidelity} to enforce data consistency by minimizing the data fidelity term $f(\mb x)$. Given $\mb v_{k-1}$ from the previous iteration, optimizing the loss $\mathcal{L}(\mb x, \mb v_{k-1})$ for $\mb x$ depends only on the data fidelity term $f(\mb x) = \frac{1}{2} \| \mathcal A(\mb x) - \mb y \|_{2}^2$ and its associated proximal term $\mu\|\mb x-\mb v_{k-1}\|_2^{2}$, leading to:
\begin{align}
\label{eqn: Pixel Space Data Fidelity}
    \mb x_k =\arg \min_{\mb x}\; \frac{1}{2} \| \mathcal A(\mb x) - \mb y \|_{2}^2+\mu \|\mb x-\mb v_{k-1}\|_2^2\:.
\end{align}
Although~\eqref{eqn: Pixel Space Data Fidelity} can be solved via iterative methods (e.g., SGD), selecting an appropriate hyperparameter $\mu$ often requires manual tuning. To mitigate this issue, we compute an \emph{approximate} solution $\mb x_k$ by minimizing only the data consistency term $f(\mb x)$, initialized at $\mb v_{k-1}$ and run for a fixed number of gradient steps. Empirically, this approach (\emph{i}) yields a solution $\mb x_k$ that stays close to $\mb v_{k-1}$ without relying on $\mu$, and (\emph{ii}) maintains low computational cost while being less sensitive to the hyperparameter.

Note that enforcing data consistency with~\eqref{eqn: Pixel Space Data Fidelity} is significantly more efficient than DPS-based methods, as it does not require backpropagation over the entire score function. In practice, we can use any gradient-based optimizer (e.g., SGD, ADAM) for minimizing~\eqref{eqn: Pixel Space Data Fidelity}. Throughout the experiments, we fix the number of gradient steps $\tau$ for each round of the data consistency optimization. When the measurements are corrupted by noise, one can reduce $\tau$ to prevent overfitting. We discuss our choice of optimizers in detail in~\Cref{choice of hyperparameters}. 
For simplicity, we denote this reconstruction phase by:
\begin{align}\label{Pixel Space Data Fidelity Simple}
    \mb x_k = \texttt{Data-Fidelity}(\mb v_{k-1}, \mb y, \mc A,\tau).
\end{align}

\begin{figure*}[t]
    \centering
    \includegraphics[width=0.8\textwidth]{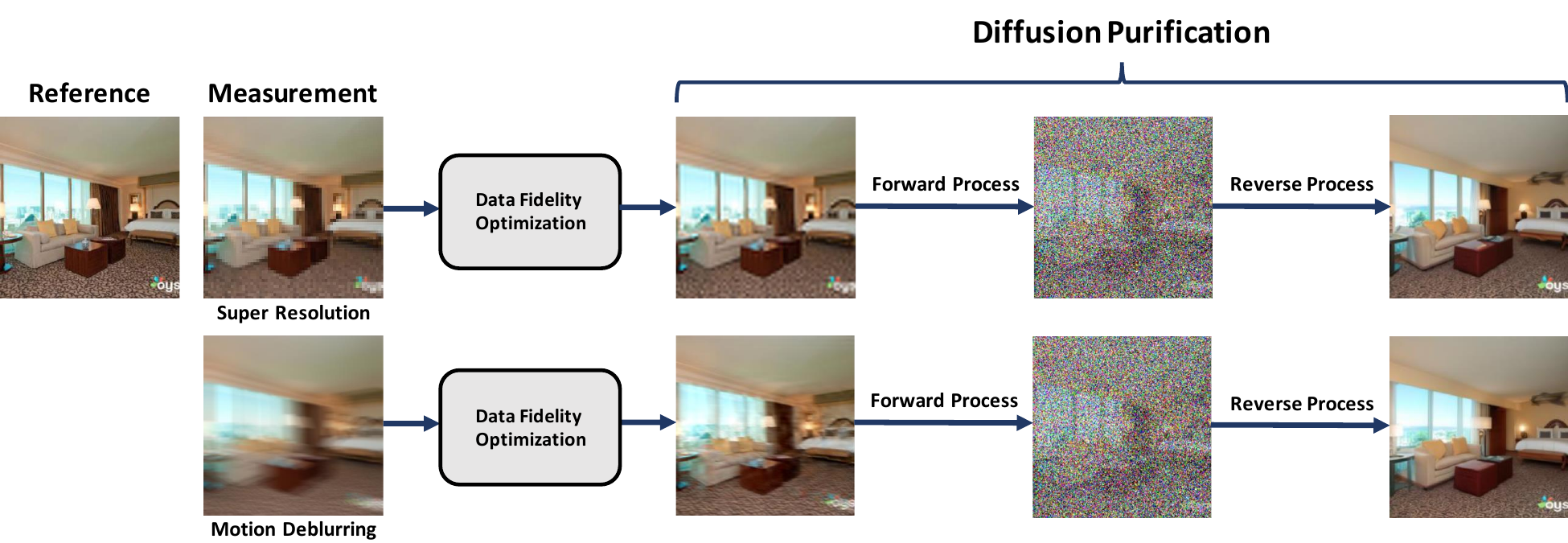}
    \caption{\textbf{Demonstrating the regularization effects of diffusion purification.} 
    The estimates from the data fidelity step contain artifacts, making the overall reconstruction inaccurate, whereas regularization via diffusion purification can remove such artifacts.}
    \label{fig: Purificaiton Effect}
\end{figure*}
\medskip
\noindent \textbf{Refinement Phase.}
Due to the ill-posed nature of the problem in \eqref{eq: forward measurement process}, $\mb x_k$ obtained in the reconstruction phase could suffer from task-specific artifacts, as shown in~\Cref{fig: Purificaiton Effect} (third column). To alleviate these artifacts from $\mb x_k$, we propose to use diffusion purification described in~\Cref{sec:dpur} as an \emph{implicit regularizer}.
Specifically, let $t_k$ denote a predefined diffusion purification timestep that controls the strength of purification at iteration $k$. Following~\eqref{eqn:diff-pur-adv}, we first define $\mb x_{k,0} := \mb x_k$ and submerge this sample with noise by running the forward process up to time $t_k$:
\begin{align}
    \mb x_{k,t_k} = \sqrt{\alpha_{t_k}}\mb x_{k,0} +\sqrt{1-\alpha_{t_k}} \mb \epsilon.
\end{align} 
Then, we run \emph{unconditional} reverse sampling via either the reverse SDE~\eqref{eq: Reverse SDE} or the reverse ODE in~\eqref{eq: Reverse ODE} starting from the initial point $\mb x_{k,t_k}$ at the time $t_k$ to obtain the purified output $\mb v_k$. As demonstrated in~\Cref{fig: Purificaiton Effect}, diffusion purification effectively removes the artifacts in the reconstructions from the data consistency (fidelity) optimization~\eqref{eqn: Pixel Space Data Fidelity} while preserving the overall image structures at the same time. For simplicity, we denote this refinement step by: 
\begin{align}\label{eq: dpur-abs}
    \mb v_k = \texttt{DPUR}(\mb x_k, \mb s_{\mb \theta}, t_k).
\end{align}
For the choice of $t_k$, we propose to use a linearly decaying schedule for robust performance, which is capable of addressing the ``hallucination'' issue brought by the stochasticity of the diffusion purification process.
In the early iterations of our method, when $t_k$ is relatively large, the resulting $\mb v_k$, though it may be inconsistent with the measurements, will be free from artifacts and can serve as a good starting point for the subsequent iteration of data consistency optimization. As $t_k$ keeps decreasing when $k$ increases, the algorithm will converge to an equilibrium point which is both consistent with the measurements and free from the artifacts. We note that, beyond the linear schedule, diffusion purification also achieves robust performance under other decaying schedules, as we will show in~\Cref{subsec:exp-ablation}. 

The overall algorithm is presented in~\Cref{alg: Pixel Space Solver}. More technical details are deferred to~\Cref{app:extra-results}, 
and we empirically study the convergence 
in~\Cref{subsec:exp-ablation}.
\begin{algorithm}[t]
    \caption{DCDP using Pixel Space Diffusion Models}\label{alg: Pixel Space Solver}
    \begin{algorithmic}
    \Require measurements $\mb y$, forward operator $\mathcal{A}(\cdot)$, pretrained diffusion model $\mb s_{\mb \theta}$, total number of iterations $K$, decaying time schedule $\{t_k \}_{k\in [K]}$ for diffusion purification, the number of data consistency iterations $\tau$, and $\mb v_0=\mb 0\in \mathbb{R}^n$
    \For{$k=1,\dots,K$}
    \State 
    $\mb x_k = \texttt{Data-Fidelity}(\mb v_{k-1}, \mb y, \mc A,\tau)$ \Comment{solve via gradient based methods for $\tau$ steps}
    \State 
    $\mb v_k = \texttt{DPUR}(\mb x_k,\mb s_{\mb \theta}, t_k)$ \Comment{Diffusion Purification}  
    \EndFor
    \State \textbf{Return} $\mb v_k$
    \end{algorithmic}
\end{algorithm}

\subsection{Fast Diffusion Purification for Refinement}
\label{subsec:fast-sampling}

In this subsection, we demonstrate two approaches to achieve faster diffusion purification: (\emph{i}) by using accelerated samplers and (\emph{ii}) by using Tweedie's formula.

\medskip
\noindent \textbf{Accelerated Samplers.}
To improve sampling efficiency, we can adapt accelerated samplers to speed up the reverse process of diffusion purification. For example, we can use DDIMs~\cite{song2020denoising}, which constructs a non-Markovian chain of the reverse process. By doing so, each round of diffusion purification can be performed with as few as $20$ unconditional reverse sampling steps (number of function evaluations, NFEs), with minimal performance degradation. We remark that existing methods such as DPS do not offer this advantage, as they typically require over $1000$ reverse sampling steps to achieve sufficient data consistency.

\medskip
\noindent \textbf{Approximation via Tweedie's Formula.}
By Tweedie's formula~\cite{robbins1992empirical,miyasawa1961empirical}, the posterior mean given noisy sample $\mb x_t$ can be estimated with the score function:
\begin{align}
    \mathbb{E}\left[\mb x_{0} \mid \mb x_{t}\right] = \frac{1}{\sqrt{\alpha_t}}(\mb x_t+(1-\alpha_t)\nabla_{\mb x_t}\log p(\mb x_t)),
\end{align}
where $\mb x_0$ denotes a sample from the ``clean'' data distribution. Equipped with this formula, one can use the posterior mean as a single-step approximation to the solution of the reverse SDE (or ODE) in diffusion purification:
\begin{align}
        \mb v_k &\approx \mathbb{E}\left[\mb x_{k,0} \mid \mb x_{k,t_k}\right] \\
        &\approx\frac{1}{\sqrt{\alpha_{t_k}}}(\mb x_{k,t_k}+(1-\alpha_{t_k})\mb s_{\mb \theta}(\mb x_{k,t_k}, t_k))\:.
\end{align}
Note that applying Tweedie's formula leads to a trade-off between computation time and reconstruction quality. The multi-step reverse SDE produces a sharp sample that approximately follows the ground truth data distribution, whereas Tweedie's formula yields the mean of all possible clean samples given the noisy sample $\mb x_t$, which is perceptually blurrier especially for large $t$. As evidenced by our experimental results in~\Cref{sec: experiments}, using Tweedie's formula may result in blurry reconstructions with lower perceptual quality. Nevertheless, DCDP with Tweedie's formula remains competitive on standard metrics and is extremely fast, requiring only one step.

\subsection{DCDP with Latent Diffusion and Consistency Models}
\label{sec: adaptation to LDMs}

By decoupling the data consistency steps from the diffusion sampling steps, our method enables the flexibility of utilizing a broad range of diffusion models in a plug-and-play fashion. To this end, we illustrate how to adapt DCDP to latent diffusion models and consistency models.


\medskip
\noindent \textbf{DCDP with Latent Diffusion Models (LDMs).}
LDMs \cite{rombach2022high} operate in a lower-dimensional latent space by utilizing an encoder--decoder pair \cite{esser2021taming} to improve the efficiency of vanilla diffusion models. Specifically, given an encoder $\mathcal{E}: \mathbb{R}^{d} \to \mathbb{R}^r$ and a decoder $\mathcal{D} : \mathbb{R}^{r} \to \mathbb{R}^d$ where $r \ll d$, LDMs reduce the computational complexity by mapping each $\mb x \in \mathbb{R}^d$ to $\mb z = \mathcal{E}(\mb x) \in \mathbb{R}^r$ and performing the entire diffusion process in the latent space. After performing the reverse steps in the latent space, the sampled latent code $\mb z_0$ is decoded to a synthesized image via $\mb x_0 = \mathcal{D}(\mb z_0)$.

Unlike previous methods~\cite{rout2024solving,song2023solving}, the flexibility of our framework allows us to (almost) seamlessly incorporate LDMs. For LDMs, we alternate between enforcing data consistency and diffusion purification, but both in the latent space:
\begin{align}
    \mb z_k &= \argmin_{\mb z} \; \mc L_{ \mathrm{LDM}}(\mb z, \wh{\mb v}_{k-1}), \label{eqn:data-const-ldm} \\
    \wh{\mb v}_k &=  \texttt{DPUR}(\mb z_k, \mb s_{\mb \phi} , t_k), \label{eqn:dpur-ldm}
\end{align}
where $ \mc L_{ \mathrm{LDM}}$ is a loss function tailored for the latent space (see \eqref{latent consistency approach 1} and \eqref{latent consistency pixel-space}), $\mb s_{\mb \phi}$ is the score function in the latent space, and $\mb z = \mathcal{E}(\mb x)$ is the latent code. Once our method converges based on specific stopping criteria, we map the converged latent code back to the pixel space through the decoder $\mc D$. The overall method is presented in \Cref{alg: LDM Solver}. In the following subsections, we discuss two approaches for enforcing data consistency.

\paragraph{\textbf{Approach I. Data Consistency in Latent Space}}
One approach for enforcing data consistency is to operate directly in the latent space by minimizing the following objective:
\begin{align}
    \label{latent consistency approach 1}
   \mb z_k = \argmin_{\mb z}\frac{1}{2}\|\mb y-\mathcal{A}(\mathcal{D}(\mb z))\|_2^2 + \mu \|\mb z-\wh{\mb v}_{k-1}\|_2^2
\end{align}
where $\wh{\mb v}_{k-1}$ denotes the output from the previous diffusion purification step. We note that this optimization objective closely resembles the method in~\cite{bora2017compressed}. As discussed in \Cref{sec:dcdp}, we approximately solve~\eqref{latent consistency approach 1} by performing a limited number of gradient-based updates on the objective $\frac{1}{2}\|\mb y - \mathcal{A}(\mathcal{D}(\mb z))\|_2^2$, initialized from $\wh{\mb v}_{k-1}$ as:
\begin{align}\label{eqn:data-ldm}
 \hspace{-0.1in}   \mb z_k = \texttt{Data-Fidelity-LDM}(\wh{\mb v}_{k-1}, \mb y, \mc A,\mc D, \tau).
\end{align}
Before diffusion purification, we run an extra step:
\begin{align}
     \mb z_k &\leftarrow \mathcal{E}(\mathcal{D}(\mb z_k)),
\end{align}
which we denote by ``Re-Encode''. The rationale behind Re-Encode is that due to the nonlinearity of the encoder $\mathcal{E}$, there could exist infinitely many solutions $\mb z$ for the optimization problem~\eqref{eqn:data-ldm}. However, since the LDMs are trained only on $\mb z = \mathcal{E}(\mb x)$, where $\mb x$ follows the image distribution, we need to first map $\mb z_k$ back to the image space with $\mb x_k = \mathcal{D}(\mb z_k)$ and then re-encode $\mb x_k$ as $\mb z_k = \mathcal{E}(\mb x_k )$.

\begin{algorithm}[t]
\caption{LDM Version of DCDP} \label{alg: LDM Solver}
\begin{algorithmic}
    \Require measurements $\mb y$, forward operator $\mathcal{A}(\cdot)$, pretrained LDM $\mb s_{\mb \phi}$, a pretrained encoder $\mathcal{E}$ and decoder $\mathcal{D}$, total number of iterations $K$, decaying time schedule $\{t_k\}_{k\in [K]}$ for diffusion purification, iteration number $\tau$, and $\wh{\mb v}_0\sim \mathcal{N}(\mb 0, \mb I)$ if enforcing data consistency in the latent space or $\mathcal{D}(\wh{\mb v}_0) =\mb 0$ if enforcing data consistency in the pixel space. 
    \For{$k=1,\dots,K$}
    \If{Approach I: enforcing data consistency in the \emph{\textbf{latent space}} }
    \State $\mb z_k = \texttt{Data-Fidelity-LDM}(\wh{\mb v}_{k-1}, \mb y, \mc A,\mc D, \tau)$ \Comment{Enforcing data consistency in the latent space}
    \State $\mb z_k \leftarrow \mathcal{E}(\mathcal{D}(\mb z_k))$\Comment{Re-Encode}
    \ElsIf{Approach II: enforcing data consistency in the \emph{\textbf{pixel space}} }
    \State $\mb x_k = \texttt{Data-Fidelity}(\mc D(\wh{\mb v}_{k-1}), \mb y, \mc A,\tau) $ \Comment{Enforcing data consistency in the pixel space}
    \State $\mb z_k = \mathcal{E}(\mb x_k)$ \Comment{Mapping to the latent space}
    \EndIf 
    \State $\wh{\mb v}_k = \texttt{DPUR}(\mb z_k, \mb s_{\mb \phi}, t_k)$ \Comment{Latent diffusion purification} 
    \State $\mb x_k = \mathcal{D}(\wh{\mb v}_k)$ \Comment{Mapping to the pixel space}
    \EndFor
\State \textbf{Return} $\mathcal{D}(\wh{\mb v}_k)$
    \end{algorithmic}
\end{algorithm}

\paragraph{\textbf{Approach II. Data Consistency in Pixel Space}}
Alternatively, we can directly enforce the data consistency \emph{in the pixel space}. Specifically, given $\wh{\mb v}_{k-1}$ as the outcome of the previous diffusion purification step, we can optimize the following objective: 
\begin{align}
    \label{latent consistency pixel-space}
   \mb x_k &= \argmin_{\mb x} \frac{1}{2}||\mb y-\mathcal{A}(\mb x)||_2^2 + \mu \|\mb x-\mathcal{D}(\wh{\mb v}_{k-1})\|_2^2.
\end{align}
Similar to that in the pixel space,~\eqref{latent consistency pixel-space} can be approximately solved by taking a limited number of gradient-based steps on $\frac{1}{2}||\mb y-\mathcal{A}(\mb x)||_2^2$ when initialized from $\mathcal{D}(\wh{\mb v}_{k-1})$, with
\begin{align}\label{eqn:data-ldm-pixel}
    \mb x_k &= \texttt{Data-Fidelity}(\mc D(\wh{\mb v}_{k-1}), \mb y, \mc A,\tau), 
\end{align}
which is equivalent to~\eqref{Pixel Space Data Fidelity Simple} with $\mb v_{k-1} = \mc D(\wh{\mb v}_{k-1})$. Following this, we run $ \mb z_k = \mathcal{E}(\mb x_k)$ to map $\mb x_k$ into the latent space for the subsequent diffusion purification step in~\eqref{eqn:dpur-ldm}.

Compared to Approach I, it is worth noting that Approach-II is much more efficient since it does not require back-propagation through the deep decoder $\mathcal{D}(\cdot)$ when computing the gradient. In practice, both methods are effective for a range of inverse problems. However, we find that Approach II is better suited for tasks like motion deblurring and Gaussian deblurring, as it consistently yields higher-quality reconstructions than Approach I (See~\Cref{fig: Latent_VS_Pixel_GD,fig: Latent_VS_Pixel_MD} in~\Cref{sec: hyperparameters}).

\medskip
\noindent \textbf{DCDP with Consistency Models.}
Recently, Song et al.~\cite{song2023consistency} proposed a class of diffusion models termed consistency models (CMs), which can effectively generate high-quality samples in a single step. 
CMs can be seamlessly incorporated into our framework. Specifically, with a pre-trained consistency model $\mb f_{\mb \theta}$, which maps a noisy sample $\mathbf x_k$ at time $t_k$ directly to its denoised estimate, the diffusion purification step in \eqref{eq: dpur-abs} can be performed in a single step as
    \begin{align}
       \mb v_k = \texttt{DPUR}(\mb f_{\mb \theta}, \mb x_k, t_k) := \mb f_{\mb \theta}(\mb x_k, t_k).
    \end{align}
Performing diffusion purification with CMs may seem similar to the approximation via Tweedie's formula, as they both require only a single NFE. However, CMs alleviate the blurriness issue brought by Tweedie's formula discussed in~\Cref{subsec:fast-sampling}. This is because unlike Tweedie's formula that estimates the posterior mean, CMs are designed specifically for sampling from the data distribution with a single step. We illustrate the effectiveness of our framework on both techniques in~\Cref{sec: exp main results}, demonstrating that one can use any model that is readily available.

\subsection{Comparison with Related Works}
\label{subsec:related-works}
Before demonstrating the effectiveness of our approach in \Cref{sec: experiments}, we first discuss how it relates to closely related recent works. For a more comprehensive overview of other diffusion model-based methods, such as \cite{wangdmplug}, we refer readers to the recent survey papers ~\cite{daras2024surveydiffusionmodelsinverse,chung2025diffusion} and to Section 2 of Zheng et al.~\cite{zhenginversebench}.

\medskip
\noindent \textbf{Plug-and-Play Diffusion Priors.} Our work, together with concurrent studies~\cite{coeurdoux2024plug,wu2024principled,xu2024provably,zhang2024improving}, belongs to a class of plug-and-play diffusion-based inverse problem solvers that alternate between data fidelity and prior steps. While all these methods leverage diffusion denoisers as implicit priors,~\cite{coeurdoux2024plug,wu2024principled,xu2024provably} directly denoise $\mb x_k$, assuming that the data fidelity step has introduced an appropriate noise level---an assumption that may not always hold. In contrast, we explicitly inject noise into $\mb x_k$ before denoising it through either iterative reverse sampling or a one-step Tweedie approximation.


This explicit noise-then-denoise process—termed diffusion purification in our work—is also implicitly adopted in DAPS~\cite{zhang2024improving}.  The main distinction lies in the data fidelity step: DAPS uses Langevin dynamics for posterior sampling, whereas our formulation in~\eqref{eqn: Pixel Space Data Fidelity} can be viewed as a maximum a posteriori (MAP) estimation. Hence, the two algorithms can be seen as complementary. We also note that DAPS achieves stronger empirical performance, likely because it uses a substantially larger number of purification steps, referred to as annealing scheduling steps in their paper, ranging from 250 to 400. In contrast, our method uses only 10 to 20 steps, as shown in Tables 6--9 in~\Cref{choice of hyperparameters}. As the two works were developed concurrently, we leave a thorough comparison of the two approaches for future work.


Lastly, the study presented by Luther et al.~\cite{luther2023ddgm} shares close ties with DCDP using Tweedie's formula in the pixel space. Their algorithm, Diffusive Denoising of Gradient-based Minimization (DDGM), is designed for solving linear tomographic reconstruction problems, which iteratively applies data consistency steps and image denoising through a pre-trained diffusion model. 
However, they only investigate the use of Tweedie's formula for regularizing the images, which often results in blurry images (see~\Cref{sec: exp main results}).
Our algorithm circumvents this issue via DDIM which performs diffusion purification with multiple diffusion reverse sampling steps. Furthermore, our framework can incorporate accelerated samplers, latent diffusion models, and consistency models. Consequently, it addresses a broader spectrum beyond tomographic reconstruction and can be readily adapted to a wide range of applications, unlike DDGM.

\medskip
\noindent \textbf{Hard Data Consistency Methods.} Another class of approaches including DiffPIR~\cite{zhang2021plug} and ReSample~\cite{song2023solving} improves upon DPS by alternating between two key steps: (\emph{i}) at each time step $t$, rather than performing a single-step data consistency update as in DPS, DiffPIR and ReSample solve a multi-step subproblem, which is known as the hard data consistency process, conditioned on the current reconstruction $\mb x_t$ to obtain an estimate $\mb x_0$ at the initial timestep, and (\emph{ii}) map $\mb x_0$ back to the diffusion trajectory at time $t-1$. While these methods may appear conceptually related to the plug-and-play approaches discussed above, hard data consistency methods generate the final reconstruction within a \textit{single} reverse diffusion process—meaning the data consistency steps of these methods remain tightly coupled with the reverse sampling. This is also true in other recent methods such as SITCOM~\cite{alkhouri2025sitcom} that integrates multiple consistencies in generating the final reconstruction in each reverse diffusion step.
In comparison, our approach employs $K$ iterations of forward and reverse diffusion processes (diffusion purification), demonstrating that data consistency steps can be fully decoupled from the reverse sampling steps. As a result, our method is compatible with one-step CMs, which are not supported by DiffPIR or ReSample.

\medskip
\noindent \textbf{Red-Diff.} RED-Diff~\cite{mardani2024a} is an algorithm that combines regularization by denoising~\cite{romano2017little} with pre-trained diffusion models within a variational Bayesian framework. It solves an optimization problem via gradient descent to minimize a data-fitting loss while maximizing the likelihood of the reconstruction under the diffusion-based denoiser prior.

\section{Experiments}\label{sec: experiments}

In this section, we demonstrate the effectiveness of our method under various settings. First, we introduce the basic experiment setup in \Cref{subsec:exp-setup}. Second, we show our main experimental results in \Cref{sec: exp main results}.
Finally, we provide an ablation study in \Cref{subsec:exp-ablation} in terms of noise robustness, computational efficiency, and convergence speed.

\subsection{Experimental Setup}\label{subsec:exp-setup}
We evaluate our method based on the following setups. The specific choices of hyperparameters are provided in~\Cref{choice of hyperparameters}.

\medskip
\noindent \textbf{Inverse Problems Used in Evaluation}
We evaluate our method on both linear and nonlinear inverse problems. For linear inverse problems, we consider the following tasks with no additive measurement noise: (i) image inpainting using a $100 \times 100$ box mask; (ii) $4\times$ image super-resolution via uniform downsampling;
(iii) Gaussian deblurring with a $61 \times 61$ Gaussian kernel of standard deviation $\sigma = 3.0$; and
(iv) motion deblurring using a $61 \times 61$ motion blur kernel with intensity $0.5$. For nonlinear inverse problems, we consider nonlinear deblurring following the experimental setup in~\cite{chung2022diffusion}, with additive Gaussian measurement noise with standard deviation $0.05$.


\begin{figure*}[t]
    \centering
    \includegraphics[width=\textwidth]{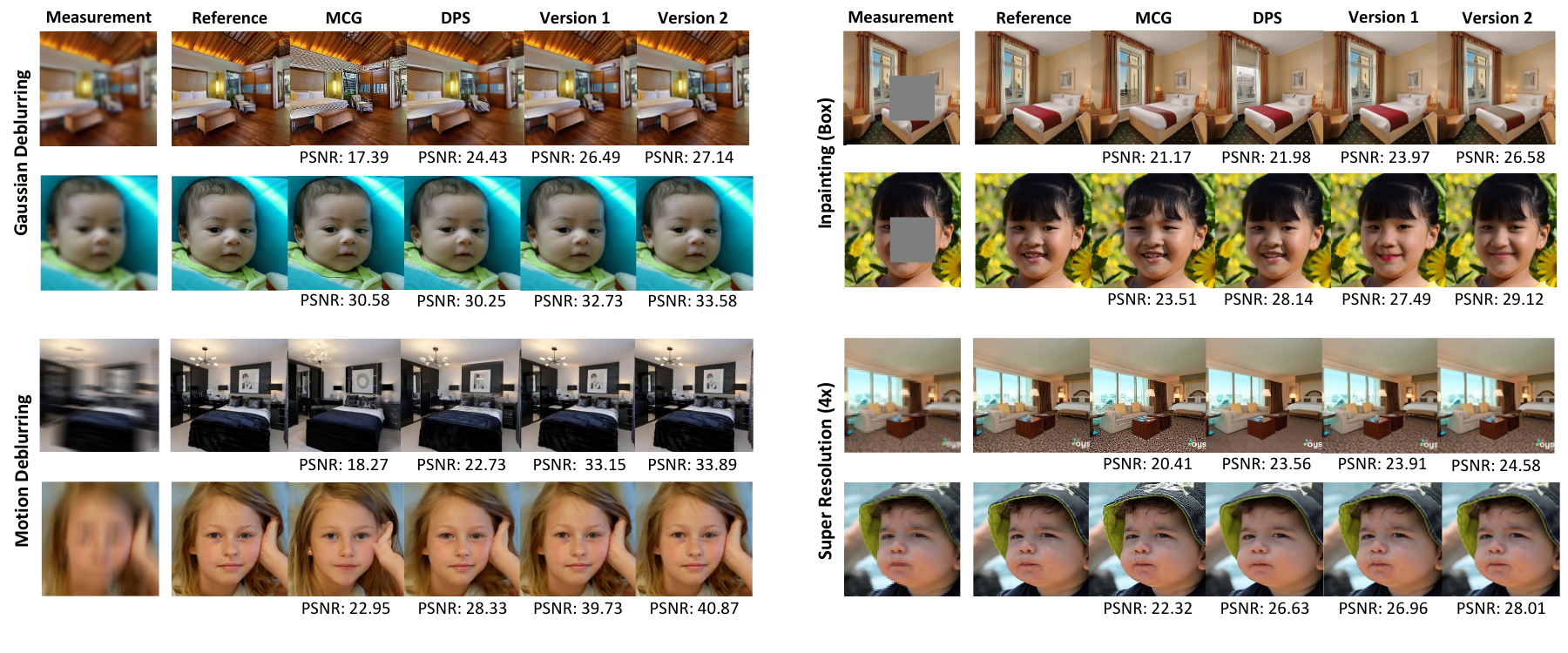}
    \caption{\textbf{Pixel space diffusion models for solving various linear inverse problems.} These results demonstrate the effectiveness of our algorithm. We can consistently recover images with the highest PSNR.}
    \label{fig: pixel space recons}
\end{figure*}

\medskip
\noindent \textbf{Evaluation Datasets.}
For linear inverse problems, we conduct experiments on the FFHQ \cite{karras2019style} and LSUN-Bedroom~\cite{yu2015lsun} datasets, both of which contain 100 images of size $256\times 256 \times 3$. For nonlinear inverse problems, we conduct experiments on ImageNet~\cite{deng2009imagenet} with 100 images of size $256\times256\times 3$.
For experiments on CMs, we only test on the LSUN-Bedroom dataset and linear inverse problems as a proof-of-concept. 

\medskip
\noindent \textbf{Evaluation Metrics.}
To quantitatively evaluate the performance, we use standard metrics: (\textit{i}) Peak Signal-to-Noise Ratio (PSNR), (\textit{ii}) Structural Similarity Index Measure (SSIM)~\cite{wang2004image} and (\textit{iii}) Learned Perceptual Image Patch Similarity (LPIPS)~\cite{zhang2018unreasonable}.

\setlength{\tabcolsep}{5pt}
\begin{table*}[t]
\begin{center}
\resizebox{1\textwidth}{!}{%
\begin{tabular}{l|lll|lll}
\hline
 \multirow{2}{*}{Method} & \multicolumn{3}{c|}{Super Resolution $4\times$} & \multicolumn{3}{c}{Inpainting (Box)} \\

& LPIPS$\downarrow$ & PSNR$\uparrow$ & SSIM$\uparrow$  & LPIPS$\downarrow$ & PSNR$\uparrow$ & SSIM$\uparrow$  \\

\hline
DPS~\cite{chung2022diffusion} & $0.225 \pm 0.05$ & $25.37 \pm 3.17$ & $0.717 \pm 0.09$ & $0.155 \pm 0.04$ & $\underline{23.58}\pm 3.49$ & $0.795 \pm 0.09$ \\
MCG~\cite{chung2022improving}  & $0.282 \pm 0.11$ & $18.94 \pm 4.34$ & $0.561 \pm 0.15$ &$0.159 \pm 0.04$ & $20.32 \pm 3.54$ & $0.789 \pm 0.06$\\
ADMM-PnP~\cite{ahmad2020plug} & $0.349 \pm 0.04$ & $21.56 \pm 2.16$ & $0.676 \pm 0.07$   & $0.232 \pm 0.02$ & $18.41 \pm 2.09$ & $0.728 \pm 0.03$ \\

DDNM~\cite{wang2022zero} & $0.233 \pm 0.06$ & $26.00 \pm 2.56$ & $0.756 \pm 0.08$   & $0.105 \pm 0.06$ & $23.42 \pm 3.45$ & $0.882 \pm 0.08$\\

RED-Diff~\cite{mardani2024a} & $0.267 \pm 0.05$ & $25.78 \pm 1.59$ & $0.746 \pm 0.07$  & $0.115 \pm 0.08$& $23.22 \pm 3.16$ & $0.878 \pm 0.05$\\

DiffPIR~\cite{zhu2023denoising} & $\underline{0.218} \pm 0.06$ & $\underline{26.34} \pm 2.34$ & $\underline{0.782} \pm 0.06$   & $0.108 \pm 0.07$& $23.47 \pm 1.89$ & $\underline{0.884} \pm 0.06$ \\

DCDP-DDIM (Ours) &$\bm{0.215} \pm 0.05$ & $26.10 \pm 3.12$& $0.768 \pm 0.08$ & $\underline{0.098} \pm 0.01$ & $20.79 \pm 2.66$  & $0.880 \pm 0.02$  \\

DCDP-Tweedie (Ours) &$0.229 \pm 0.05$ & $\bm{26.92} \pm 3.33 $& $\bm{0.800} \pm 0.07$ & $\bm{0.094} \pm 0.02$ & $\bm{23.66} \pm 3.42$  & $\bm{0.906} \pm 0.02$  \\

\hline
\rowcolor{Columnbia Blue}
 Latent-DPS &  $0.415 \pm 0.06$& $22.96 \pm 2.62$  & $0.573 \pm 0.11$  & $0.312 \pm 0.05$ & $18.66 \pm 2.53$ & $0.656 \pm 0.08$\\
 \rowcolor{Columnbia Blue}
  PSLD~\cite{rout2024solving}  & $0.353 \pm 0.07$  & $24.53 \pm 2.66$  & $0.694 \pm 0.10$   &$0.336 \pm 0.07$ & $22.01 \pm 3.15$ & $0.708 \pm 0.11 $ \\

\rowcolor{Columnbia Blue}
  ReSample~\cite{song2023solving} & $\bm{0.187} \pm 0.04$ & $25.92 \pm 3.20$  & $0.764 \pm 0.08$ & $\bm{0.195} \pm 0.04$ & $21.05 \pm 3.11$ & $0.762 \pm 0.07$\\
\rowcolor{Columnbia Blue}
  DCDP-LDM-DDIM (Ours) &$\underline{0.244} \pm 0.05$ & $\underline{26.52} \pm 3.21$& $\underline{0.778} \pm 0.08$ & $\underline{0.215} \pm 0.04$ & $\underline{23.75} \pm 3.32$  & $\bm{0.831} \pm 0.05$  \\
\rowcolor{Columnbia Blue}
  DCDP-LDM-Tweedie (Ours) &$0.259 \pm 0.05$ & $\bm{26.88} \pm 3.28$& $\bm{0.789} \pm 0.08$ & $0.239 \pm 0.04$ & $\bm{24.22} \pm 3.65$  & $\underline{0.828} \pm 0.05$  \\
\hline
\rowcolor{orange}
DCDP-CM (Ours)
&$0.208 \pm 0.05$ & $26.17 \pm 3.19$& $0.774 \pm 0.08$ 
& $0.108 \pm 0.02$ & $19.66 \pm 2.48$  & $0.874 \pm 0.02$\\
\hline
\end{tabular}}
\end{center}
\caption{\textbf{Quantitative results of super resolution and inpainting on the LSUN-Bedroom dataset for both pixel space diffusion models and LDMs.} The results colored in \textcolor{Columnbia Blue}{blue} and in \textcolor{orange}{orange} are for LDMs and CMs respectively, while the others are for pixel space diffusion models. The best results are in bold and second best results are underlined for pixel space models and LDMs,  respectively.}

\label{table: quant1}
\end{table*}

\medskip
\noindent \textbf{Comparison Baselines.}
We conduct experiments using both pixel space and latent space diffusion models. For the pixel space, we compare our algorithm with DPS~\cite{chung2022diffusion}, Manifold Constrained Gradients (MCG)~\cite{chung2022improving}, Plug-and-Play ADMM~\cite{ahmad2020plug} (PnP-ADMM), DiffPIR~\cite{zhu2023denoising}, RED-Diff~\cite{mardani2024a} and DDNM~\cite{wang2022zero}.
For the latent space, we consider Latent-DPS~\cite{song2023solving}, ReSample~\cite{song2023solving}, and Posterior Sampling with Latent Diffusion Models (PSLD)~\cite{rout2024solving}. To ensure a fair comparison, both the baselines and DCDP (ours) use the same sets of pre-trained diffusion models, including pixel space diffusion models from~\cite{chung2022diffusion} and~\cite{dhariwal2021diffusion}, LDMs from~\cite{rombach2022high}, and CMs from~\cite{song2023consistency}. For both DCDP and the baselines considered, we tune the hyperparameters via grid search to obtain the best possible performance. 


\setlength{\tabcolsep}{5pt}
\begin{table*}[t]
\begin{center}
\resizebox{\textwidth}{!}{%
\begin{tabular}{l|lll|lll}
\hline
 \multirow{2}{*}{Method} & \multicolumn{3}{c|}{Gaussian Deblurring} & \multicolumn{3}{c}{Motion Deblurring}  \\

& LPIPS$\downarrow$ & PSNR$\uparrow$ & SSIM$\uparrow$  & LPIPS$\downarrow$ & PSNR$\uparrow$ & SSIM$\uparrow$ \\

\hline
 DPS~\cite{chung2022diffusion} & $0.213 \pm 0.08$ & $24.45 \pm 3.72$ & $0.691 \pm 0.132$ & $0.182 \pm 0.03$ &  $24.45 \pm 2.93$ & $0.736 \pm 0.06$  \\

  MCG~\cite{chung2022improving} & $0.311 \pm 0.13$ & $17.54 \pm 5.06$ & $0.551 \pm 0.191$ & $0.365 \pm 0.11$ &  $20.17 \pm 3.73$ & $0.515 \pm 0.36$  \\

  ADMM-PnP~\cite{ahmad2020plug} &$0.437 \pm 0.04$ & $20.76 \pm 1.94$& $0.595 \pm 0.09$ & $0.524 \pm 0.04$ & $18.05 \pm 2.05$  & $0.493 \pm 0.10$  \\
DDNM~\cite{wang2022zero} & $0.217\pm 0.07$ & $\underline{27.30} \pm 1.67$ &  $\underline{0.815} \pm 0.08$   & --& --&--\\

RED-Diff~\cite{mardani2024a} & $0.228 \pm 0.12$ & $23.98 \pm 1.23$ & $0.687 \pm 0.067$ & $0.178 \pm 0.06$  & $24.15 \pm 2.34$ & $0.704 \pm 0.12$\\

DiffPIR~\cite{zhu2023denoising} & $0.214 \pm 0.013$ & $27.04 \pm 2.67$ & $0.801 \pm 0.12$   & $0.114 \pm 0.12$ & $27.45 \pm 1.89$ & $0.866 \pm 0.07$\\
  DCDP-DDIM (Ours) &$\bm{0.196} \pm 0.05$ & $27.13 \pm 3.26$& $0.804 \pm 0.07$ & $\underline{0.065} \pm 0.01$ & $\underline{33.56} \pm 3.92$  & $\underline{0.947} \pm 0.02$  
  \\
  
  DCDP-Tweedie (Ours) &$\underline{0.212} \pm 0.05$ & $\bm{27.74} \pm 3.34$& $\bm{0.825} \pm 0.07$ & $\bm{0.067} \pm 0.02$ & $\bm{34.47} \pm 4.07$  & $\bm{0.956} \pm 0.02$  \\


\hline
\rowcolor{Columnbia Blue}
  Latent-DPS &  $0.337 \pm 0.05$& $23.75 \pm 2.53$  & $0.622 \pm 0.10$  & $0.425 \pm 0.06$ & $21.90 \pm 2.31$ & $0.539 \pm 0.10$ \\
  \rowcolor{Columnbia Blue}
  PSLD~\cite{rout2024solving}  & $0.373 \pm 0.07$ & $24.26 \pm 2.84$  & $0.683 \pm 0.11$  & $0.469 \pm 0.06$ & $20.58 \pm 2.32$ & $0.562 \pm 0.11$ \\
  \rowcolor{Columnbia Blue}
  ReSample~\cite{song2023solving}  & $\underline{0.240} \pm 0.05$ & $25.76 \pm 3.02$  & $0.731 \pm 0.09$ & $0.188 \pm 0.04$ & $27.96 \pm 3.07$ & $0.806 \pm 0.07$ \\
  \rowcolor{Columnbia Blue}
  DCDP-LDM-DDIM (Ours) &$0.246 \pm 0.05$ & $\underline{26.08} \pm 3.02$ & $\underline{0.766} \pm 0.07$ & $\underline{0.145} \pm 0.03$ & $\underline{28.87} \pm 3.77$  & $\underline{0.856} \pm 0.06$ \\ 
\rowcolor{Columnbia Blue}
  DCDP-LDM-Tweedie (Ours) &$\bm{0.217} \pm 0.05$ & $\bm{27.11} \pm 3.26$& $\bm{0.807} \pm 0.07$ & $\bm{0.121} \pm 0.03$ & $\bm{29.40} \pm 3.81$  & $\bm{0.875} \pm 0.05$  \\
\hline
\rowcolor{orange}
DCDP-CMs (Ours)
&$0.202 \pm 0.05$ & $26.91 \pm 3.11$& $0.805 \pm 0.07$ 
& $0.061 \pm 0.02$ & $34.10 \pm 3.80$  & $0.953 \pm 0.02$\\
\hline
\end{tabular}}
\end{center}
\caption{\textbf{Quantitative results of Gaussian and motion deblurring on the LSUN-Bedroom dataset for both pixel space diffusion models and LDMs.} The results colored in \textcolor{Columnbia Blue}{blue} and \textcolor{orange}{orange} are for LDMs and CMs respectively, while the others are for pixel space diffusion models. The best results are in bold and second best results are underlined for pixel space models and LDMs respectively.}
\label{table: quant2}
\end{table*}

\subsection{Experimental Results}
\label{sec: exp main results}
In this section we present both quantitative and qualitative results for our proposed algorithm and the baselines. For conciseness, we defer the quantitative results for FFHQ to~\Cref{app:extra-results}. As discussed in~\Cref{subsec:fast-sampling}, the diffusion purification can be performed by either (i) employing accelerated samplers such as DDIM or (ii) applying a single-step approximation using Tweedie's formula. We denote these variants as DCDP-DDIM and DCDP-Tweedie, respectively—and as DCDP-LDM-DDIM and DCDP-LDM-Tweedie when implemented with LDMs—and evaluate both of them in our experiments.



\medskip 
\begin{wrapfigure}{R}{0.46\linewidth}
    \centering
    \includegraphics[width=\linewidth]{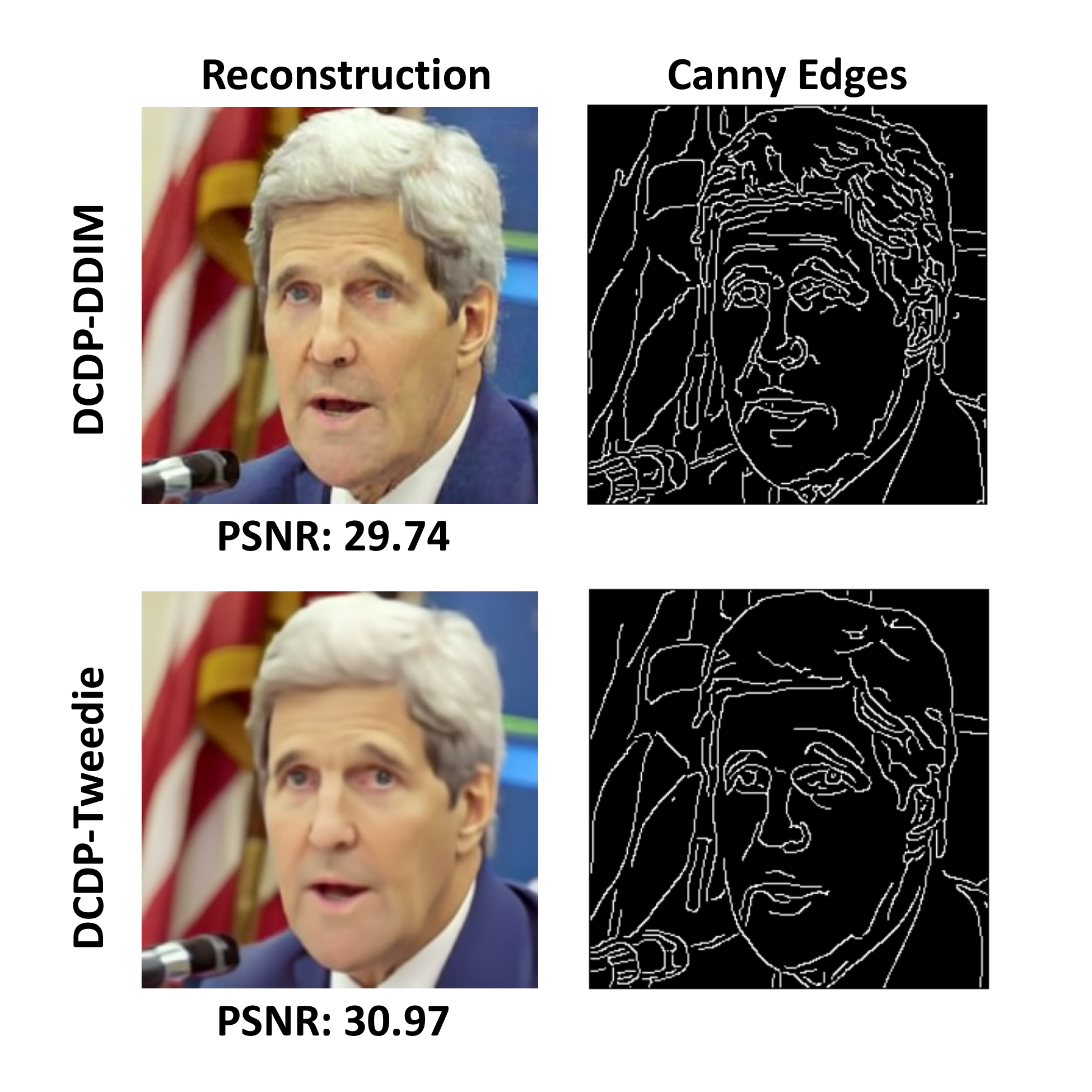}
    \caption{\textbf{Demonstrating the differences between the DCDP methods on SR ($4\times$).} The two columns show the reconstructions with PSNRs and their corresponding Canny edges.}
    \label{fig: canny edges}
\end{wrapfigure}
\noindent \textbf{Comparison of Pixel Space Diffusion Models.}\par
We first test DCDP and the baselines based on pixel space diffusion models. The qualitative results are reported in~\Cref{fig: pixel space recons}, and the quantitative results are presented in~\Cref{table: quant1,table: quant2,table: quant nonlinear deblur}. In these tables, it is evident that our methods consistently outperform or achieve performance on par with the baselines. We note that Red-Diff~\cite{mardani2024a} performs extremely well on nonlinear deblurring but is less competitive on other tasks.

When comparing DCDP-DDIM and DCDP-Tweedie, we observe that DCDP-Tweedie generally achieves higher PSNR and SSIM, which mainly measure the data fidelity. However, its reconstructions are often blurrier and contain less perceptual detail. As shown in~\Cref{fig: canny edges}, DCDP-DDIM qualitatively preserves richer high-frequency content—evident in the Canny-edge visualizations—and leads to a better LPIPS score, which better aligns with the perceptual quality. This phenomenon is commonly referred to as the ``perception-distortion trade-off''~\cite{blau2018perception}. 

\begin{figure*}[t]
    \centering
    \includegraphics[width=0.9\textwidth]{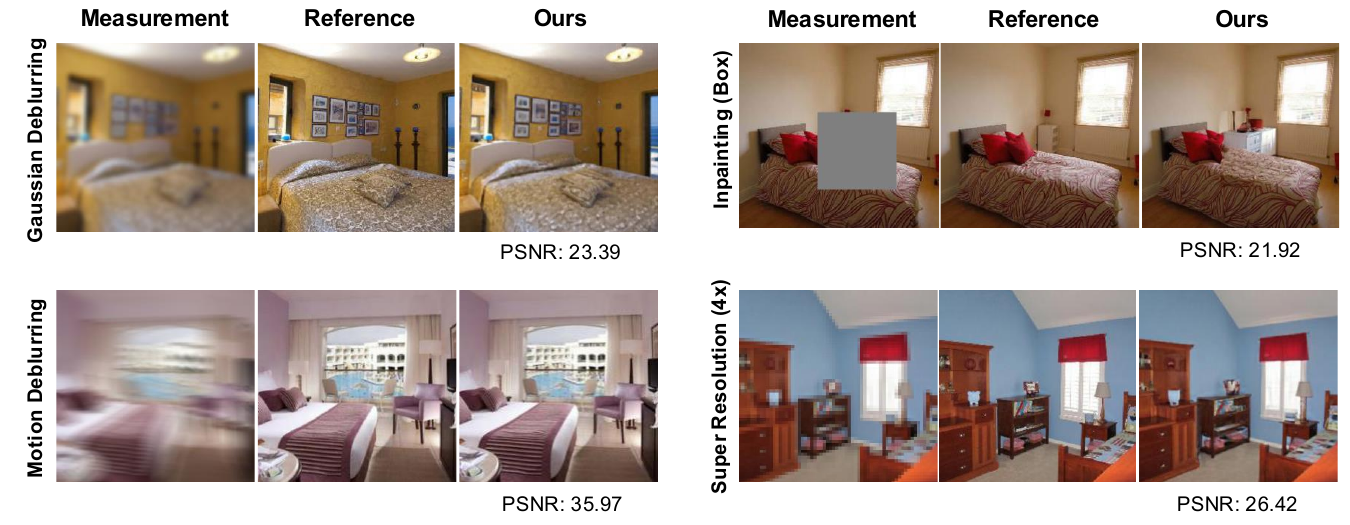}
    \caption{\textbf{Qualitative results for various image restoration problems by using CMs.} Our proposed framework is capable of utilizing CMs for image restoration.}
    \label{fig: CMs Result}
\end{figure*}

For the results with CMs, we present the quantitative results in~\Cref{table: quant1,table: quant2} highlighted in orange, and qualitative results in \Cref{fig: CMs Result}, which demonstrate that DCDP with CMs leads to decent reconstruction quality for the inverse problems considered. This result highlights the flexibility of our method to accommodate a broad range of diffusion-based models. In contrast, none of the baselines considered are able to utilize CMs, as they require a sufficiently large number of reverse sampling steps. Lastly, we note that the CMs used in this work are adopted from ~\cite{song2023consistency}, which is the first-ever prototype that introduces the concept of CMs. The performance of DCDP with CMs could potentially be further improved by incorporating more advanced consistency models proposed in recent works such as sCMs~\cite{lu2024simplifying}, Latent Consistency Models~\cite{luo2023latent} and InstaFlow~\cite{liu2023instaflow}, which we leave for future investigation.

\begin{figure*}[t]
    \centering    
    \includegraphics[width=\textwidth]{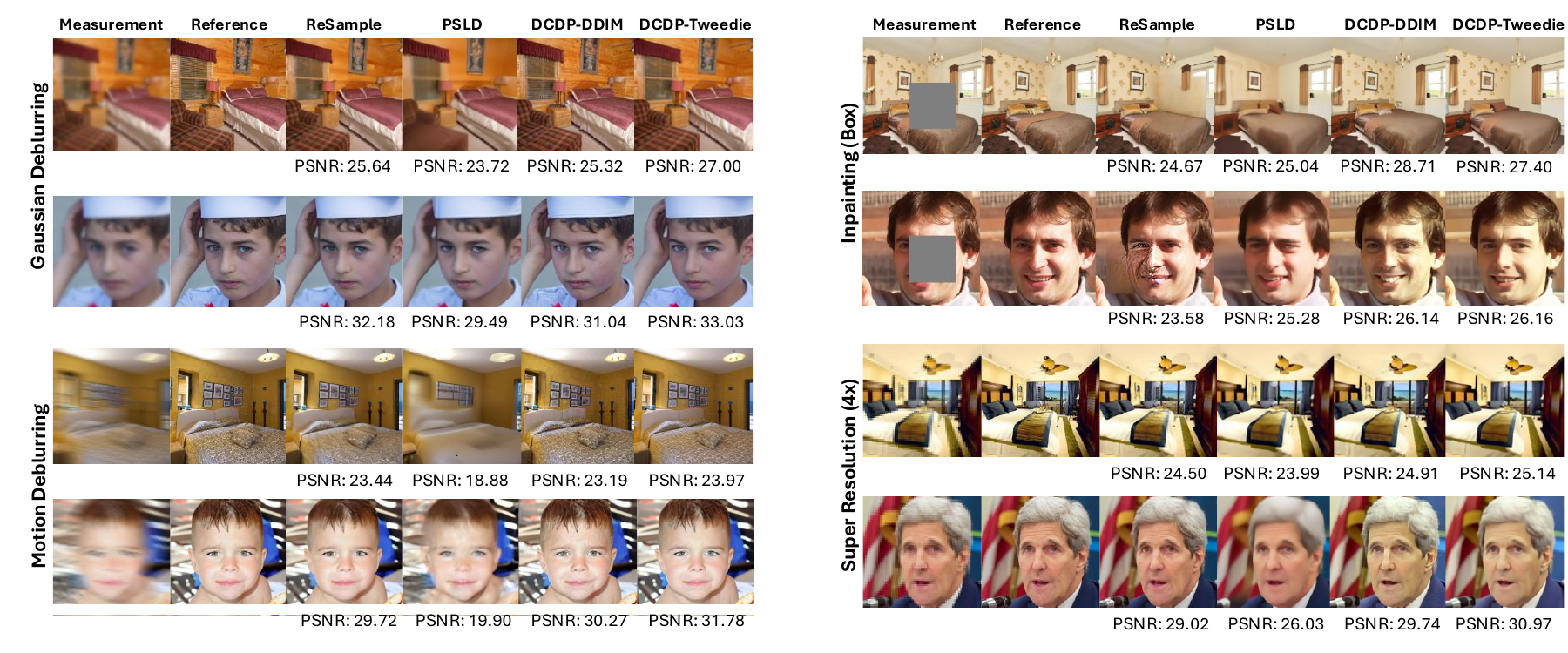}
    \caption{\textbf{Latent space diffusion models for solving various image restoration problems.} These results demonstrate the effectiveness of our algorithm. We can consistently recover images with the highest PSNR.}
    \label{fig: latent recons}
\end{figure*}
\medskip 
\noindent \textbf{Comparison of Latent Space Diffusion Models.}
For the results with LDMs, we present quantitative results in~\Cref{table: quant1,table: quant2,table: quant nonlinear deblur} highlighted in blue, and qualitative results in~\Cref{fig: latent recons}. Recall that, as outlined in~\Cref{sec: adaptation to LDMs}, there are two approaches for enforcing data consistency. We use pixel space data consistency (Approach II) for motion deblurring and Gaussian deblurring, and latent space optimization (Approach I) for box inpainting, super-resolution, and nonlinear deblurring, as doing so empirically leads to the best reconstruction quality. As shown in the tables, our algorithm consistently matches or outperforms all existing baselines. Interestingly, we observe that PSLD seems to reconstruct images that are often noisy and blurry, where the learning rate plays a huge role in the reconstruction quality. 
While ReSample could achieve lower (better) LPIPS scores on some inverse problem tasks, our algorithm is able to achieve better PSNR and SSIM scores. 

\setlength{\tabcolsep}{5pt}
\begin{table}[t]
\begin{center}
\resizebox{0.8\textwidth}{!}{%
\begin{tabular}{l|lll}
\hline
 \multirow{2}{*}{Method} & \multicolumn{3}{c}{Nonlinear Deblurring}   \\

& LPIPS$\downarrow$ & PSNR$\uparrow$ & SSIM$\uparrow$\\

\hline
 DPS~\cite{chung2022diffusion} & $0.355 \pm 0.09$& $22.60 \pm 3.36$ & $0.588 \pm 0.16$ \\
RED-Diff~\cite{mardani2024a} & $\textbf{0.211}\pm 0.05$ & $\textbf{29.51}\pm 0.76$ & $\textbf{0.828}\pm 0.08$  \\
  DCDP-DDIM (Ours) &$0.323\pm 0.06$& $24.08\pm 2.42$ & $0.634\pm 0.112$
  \\
  
  DCDP-Tweedie (Ours) &$\underline{0.320}\pm0.08$& $\underline{25.36}\pm 2.77$& $\underline{0.712}\pm 0.10$ \\


\hline
\rowcolor{Columnbia Blue}
  Latent-DPS &  $0.483 \pm 0.08$ & $21.77 \pm 3.02$  & $0.513 \pm 0.16$ \\
  \rowcolor{Columnbia Blue}
  ReSample~\cite{song2023solving}  & $\bm{0.279} \pm 0.07$ & $\textbf{26.36} \pm 2.90$  & $0.717 \pm 0.11$ \\
  \rowcolor{Columnbia Blue}
  DCDP-LDM-DDIM (Ours) &$\underline{0.286} \pm 0.08$ & $25.97 \pm 3.23$ & $\underline{0.735} \pm 0.12$ \\ 
\rowcolor{Columnbia Blue}
  DCDP-LDM-Tweedie (Ours)  & $0.302 \pm 0.09$ & $\underline{26.34} \pm 3.34$  & $\textbf{0.747} \pm 0.12$ \\
\hline
\end{tabular}}
\end{center}
\caption{\textbf{Quantitative results of nonlinear deblurring on the ImageNet dataset for both pixel space diffusion models and LDMs.} The results colored in \textcolor{Columnbia Blue}{blue} are for LDMs, while the others are for pixel space diffusion models. The best results are in bold and second best results are underlined for pixel space models and LDMs respectively.}
\label{table: quant nonlinear deblur}
\end{table}

\subsection{Ablation Studies}
\label{subsec:exp-ablation}

Finally, we conduct ablation studies demonstrating computational efficiency, convergence, and robustness of the proposed DCDP method.

\Needspace{0.72\textheight}
\medskip
\begin{wrapfigure}{r}{0.5\linewidth}
    \centering
    \includegraphics[width=\linewidth]{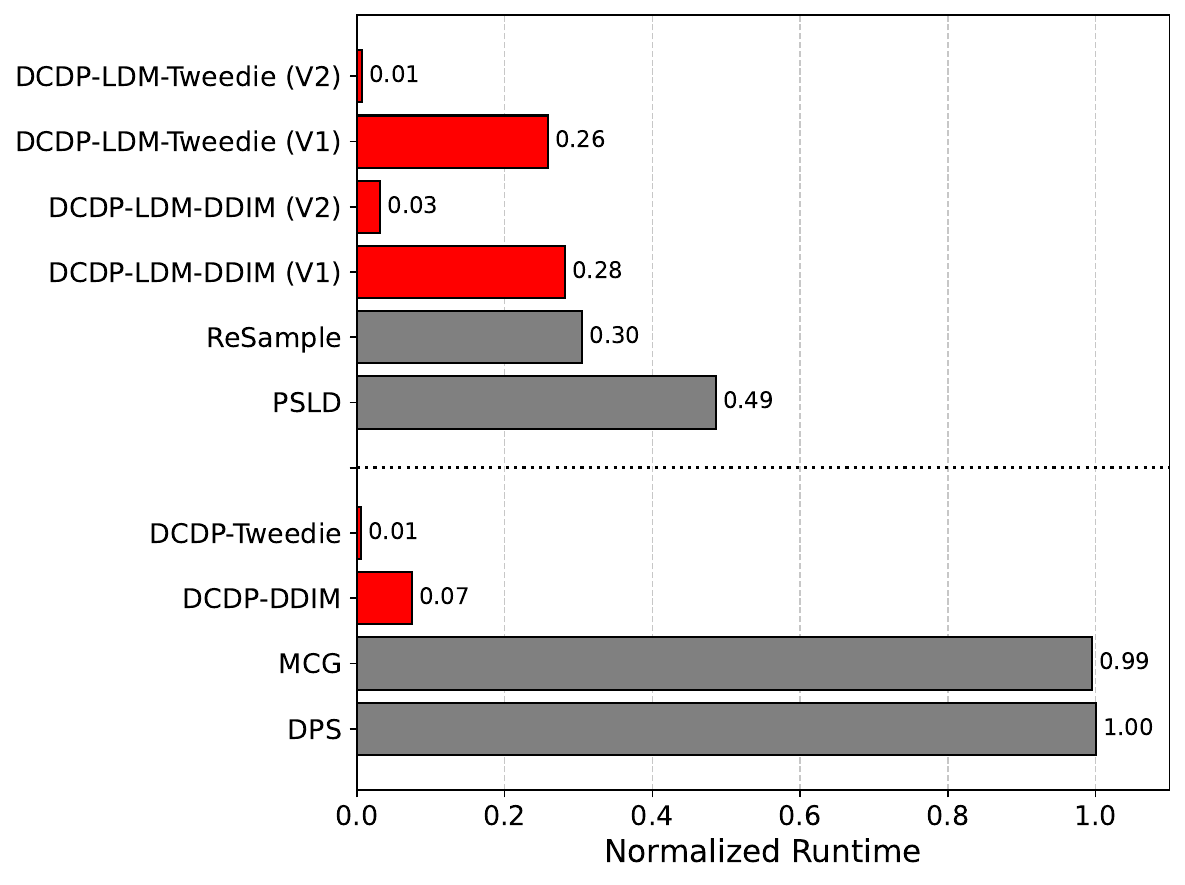}
    \caption{\textbf{Comparison of inference times on Motion Deblurring.} The experiment setting is the same as that introduced in~\Cref{subsec:exp-setup} and all methods are run using the exact same compute. Here V1 and V2 stand for Approach I and Approach II introduced in~\Cref{sec: adaptation to LDMs}, respectively. The inference time for our algorithm is the fastest, while enjoying similar (or better) performance. The runtime is normalized by dividing by the largest value, which corresponds to DPS with a total runtime of 387 seconds. }
    \label{fig:enter-label}
\end{wrapfigure}
\noindent \textbf{Computational Efficiency.} First, we compare the inference speed (the time it takes to reconstruct a single image) of our method with the baselines and display the results in~\Cref{fig:enter-label}. Clearly, for pixel-based diffusion models, our methods are much more efficient than DPS~\cite{chung2022diffusion} and MCG \cite{chung2022improving}, as we do not require backpropagation through the score function for data consistency. Furthermore, using the single-step Tweedie's formula approximation for diffusion purification (DCDP-Tweedie and DCDP-LDM-Tweedie) is much faster than using the reverse ODE (DDIM) which involves multiple reverse diffusion steps (DCDP-DDIM and DCDP-LDM-DDIM). Second, for LDMs, our methods are faster than both ReSample and PSLD. Moreover, enforcing data consistency in the pixel space (Approach II) is considerably more efficient than enforcing data consistency in the latent space (Approach I), as the former does not require backpropagation through the deep decoder $\mc D(\cdot)$. 

\begin{figure*}[t]
    \centering
    \includegraphics[width=\textwidth]{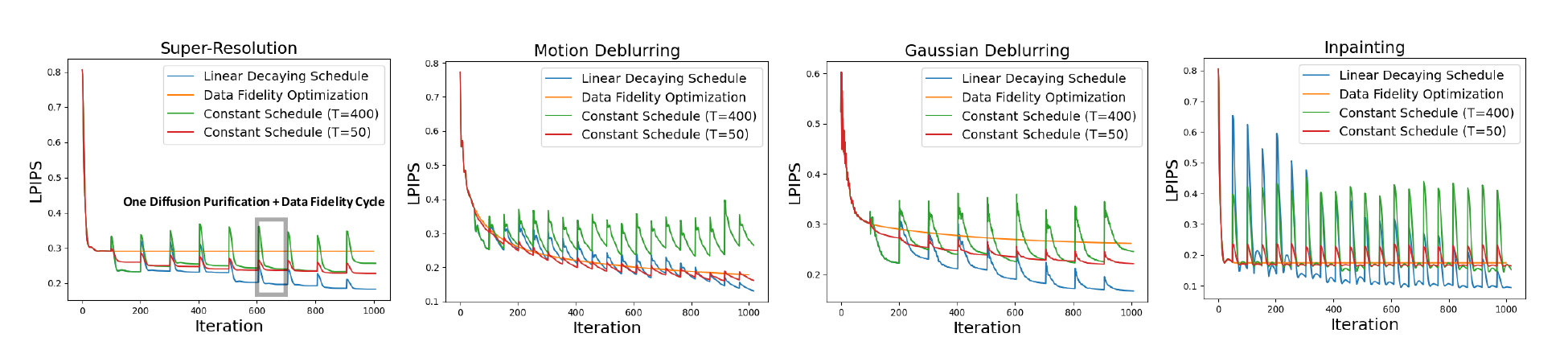}
    \caption{\textbf{Convergence of our methods in terms of LPIPS.} The orange, green, red and blue curves correspond to the reconstruction process with data fidelity optimization only, DCDP with a constant (large) purification strength, constant (small) purification strength and a linearly decaying purification strength, respectively. The x-axis denotes the total number of data fidelity gradient steps. See~\Cref{fig:dcdp convergence behavior nonlinear deblur.} in~\Cref{app:extra-results} for convergence behavior measured using PSNR and SSIM.}
    \label{fig: Convergence}
\end{figure*}

\begin{figure*}[t]
         \centering
         \includegraphics[width=0.9\textwidth]{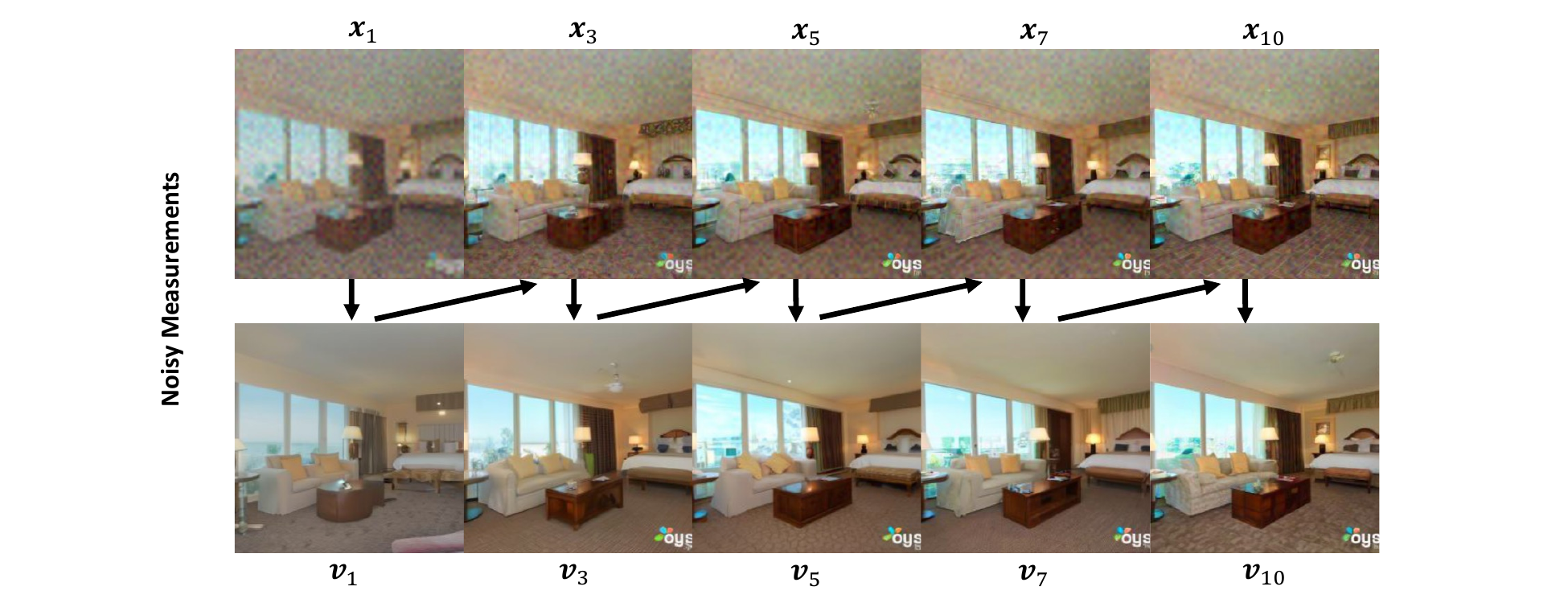}
         \caption{\textbf{Reconstruction process for super-resolution (4x) in the pixel space.} To compensate the artifacts introduced by the measurement noise, the reconstruction process takes $K=10$ iterations in total with a time schedule starting at $t_1=400$ and linearly decaying to $t_{10}=200$. The learning rate for data fidelity optimization is $lr=100$.  
         }
         \label{fig: Process Noisy}
\end{figure*}

\medskip 
\noindent \textbf{Empirical Convergence Study.} 
We further investigate the convergence of our algorithm in terms of LPIPS. This study also serves to demonstrate the efficiency of a linearly decaying time schedule for $t_k$ in diffusion purification.
As shown in~\Cref{fig: Convergence}, at the beginning of each diffusion purification and data fidelity optimization cycle, the LPIPS increases after performing diffusion purification due to stochasticity introduced by the forward-backward diffusion process. However, the subsequent iterations successfully decrease LPIPS and result in higher-quality reconstructions than those obtained using pure data fidelity optimization, implying that the purified image is able to serve as a good initialization point. In the same figure, note that the linearly decaying schedule greatly improves the performance of our algorithm compared to using a constant time schedule for $t_k$. This is expected, as a large constant purification strength will result in serious hallucination while a small constant purification strength is not able to fully alleviate the artifacts in the data-fidelity-based estimations. We note that similar convergence behavior holds when measuring the reconstruction quality with PSNR and SSIM, as shown in~\Cref{fig:dcdp convergence behavior nonlinear deblur.} in~\Cref{app:extra-results}. Lastly, in~\Cref{fig: Process Noisy}, we illustrate an example reconstruction process for super-resolution with noisy measurements, with more qualitative examples in the Appendix. It is evident that the quality of the reconstructed image gradually increases with each iteration, implying that the diffusion purification is an effective operator for enforcing the image priors embedded in the pre-trained diffusion models.

\begin{table*}[t]
\begin{center}
\resizebox{1\textwidth}{!}{%
\begin{tabular}{{ll|lll|lll}}
\hline
&  & \multicolumn{3}{c|}{Super Resolution ($4\times$)} & \multicolumn{3}{c}{Motion Deblurring}\\
  \multicolumn{2}{c|} {Metric}&DPS & DCDP-DDIM & DCDP-Tweedie & DPS & DCDP-DDIM & DCDP-Tweedie \\
\hline
\multirow{3}{*}{($\sigma_{\bm y} = 0.050$)}&LPIPS$\downarrow$ & $\mathbf{0.287} \pm 0.03$ & $\underline{0.321} \pm 0.05$ & $0.363 \pm 0.05$ & $\mathbf{0.328} \pm 0.07$ & $\underline{0.332} \pm 0.05$ & $0.361 \pm 0.05$\\ 
&PSNR$\uparrow$ & $\underline{24.50} \pm 2.88$ & $24.23 \pm 2.45$ & $\mathbf{24.66} \pm 2.55$ & $22.43 \pm 2.83$ & $\underline{23.47} \pm 2.23$ & $\mathbf{24.18} \pm 2.31$ \\

&SSIM$\uparrow$& $\underline{0.697} \pm 0.08$ & $0.687 \pm 0.09$ & $\mathbf{0.709} \pm 0.09$ & $0.642 \pm 0.09$ & $\underline{0.669} \pm 0.09$ & $\mathbf{0.703} \pm 0.09$ \\
\hline
\multirow{3}{*}{($\sigma_{\bm y} = 0.075$)}&LPIPS$\downarrow$ & $\underline{0.342}\pm 0.04$ & $\mathbf{0.325} \pm 0.05$ & $0.363 \pm 0.05$ & $\underline{0.343} \pm 0.07$ & $\mathbf{0.340} \pm 0.05$ & $0.357 \pm 0.05$\\ 
&PSNR$\uparrow$ & $23.72 \pm 2.59$ & $\underline{24.12} \pm 2.41$ & $\mathbf{24.62} \pm 2.54$ & $22.09 \pm 2.77$ & $\underline{23.17} \pm 2.20$ & $\mathbf{24.13} \pm 2.29$ \\

&SSIM$\uparrow$& $0.644 \pm 0.08$ & $\underline{0.681} \pm 0.09$ & $\mathbf{0.709} \pm 0.09$ & $0.623 \pm 0.11$ & $\underline{0.649} \pm 0.09$ & $\mathbf{0.699} \pm 0.09$ \\
\hline
\multirow{3}{*}{($\sigma_{\bm y} = 0.100$)}&LPIPS$\downarrow$ & $0.393 \pm 0.04$ & $\mathbf{0.339} \pm 0.05$ & $\underline{0.362} \pm 0.05$ & $\mathbf{0.352} \pm 0.07$ & $0.378 \pm 0.05$ & $\underline{0.368} \pm 0.05$\\ 
&PSNR$\uparrow$ & $22.85 \pm 2.39$ & $\underline{23.92} \pm 2.34$ & $\mathbf{24.57} \pm 2.52$ & $21.76 \pm 2.70$ & $\underline{22.49} \pm 2.06$ & $\mathbf{24.05} \pm 2.26$ \\

&SSIM$\uparrow$& $0.599 \pm 0.11$ & $\underline{0.668} \pm 0.08$ & $\mathbf{0.707} \pm 0.09$ & $\underline{0.614} \pm 0.11$ & $0.593 \pm 0.09$ & $\mathbf{0.690} \pm 0.09$ \\
\hline
\end{tabular}}
\end{center}
\caption{\textbf{Case study of noise robustness of our methods compared to DPS on two image restoration problems in the pixel space.} We compare our algorithm to DPS on super resolution ($4 \times$) and motion deblurring to demonstrate that our method is effective even in the presence of measurement noise. Best results are bolded and second best results are underlined.  
}
\label{table: combined case study}

\end{table*}
\medskip 
\noindent \textbf{Noise Robustness.} 
In many practical settings of interest, the measurements $\mb{y}$ can be corrupted with additive noise. Here, we perform an ablation study on the performance of our algorithm in which the measurements have additive Gaussian noise $\mathcal{N}(0, \sigma^2_{\mb y}\mb{I})$ with varying values of $\sigma_{\mb y}$ for super-resolution and motion deblurring. We compare our algorithm to DPS, as their theory states that the approximation of the likelihood gradient becomes tighter as the measurement noise $\sigma_{\mb y}$ increases~\cite{chung2022diffusion}. 
We conduct experiments on $100$ randomly selected images from the LSUN-Bedroom dataset for super-resolution, motion deblurring with Gaussian measurement noise of standard deviations $0.05, 0.075$, and $0.1$. For the hyperparameters, we tune the parameters according to $\sigma_{\bm{y}} = 0.1$ and fix them throughout the experiments. The rest of the setup is the same as previously described. We present the results in \Cref{table: combined case study}. 

Note that as the standard deviation increases, the performance of all algorithms degrades. However, our algorithm consistently outperforms DPS in metrics such as PSNR and SSIM, whereas DPS often exhibits higher perceptual quality measured by LPIPS. This difference arises because DPS frequently suffers from hallucinations, introducing artifacts in the image that ``sharpen'' it. Our algorithm is less susceptible to such artifacts, resulting in ``smoother'' images with higher signal-to-noise ratios. Details of the hyperparameters and visual results can be found in~\Cref{sec: robust study detail}. 

\begin{figure*}[t]
\centering
\includegraphics[width=0.8\textwidth]{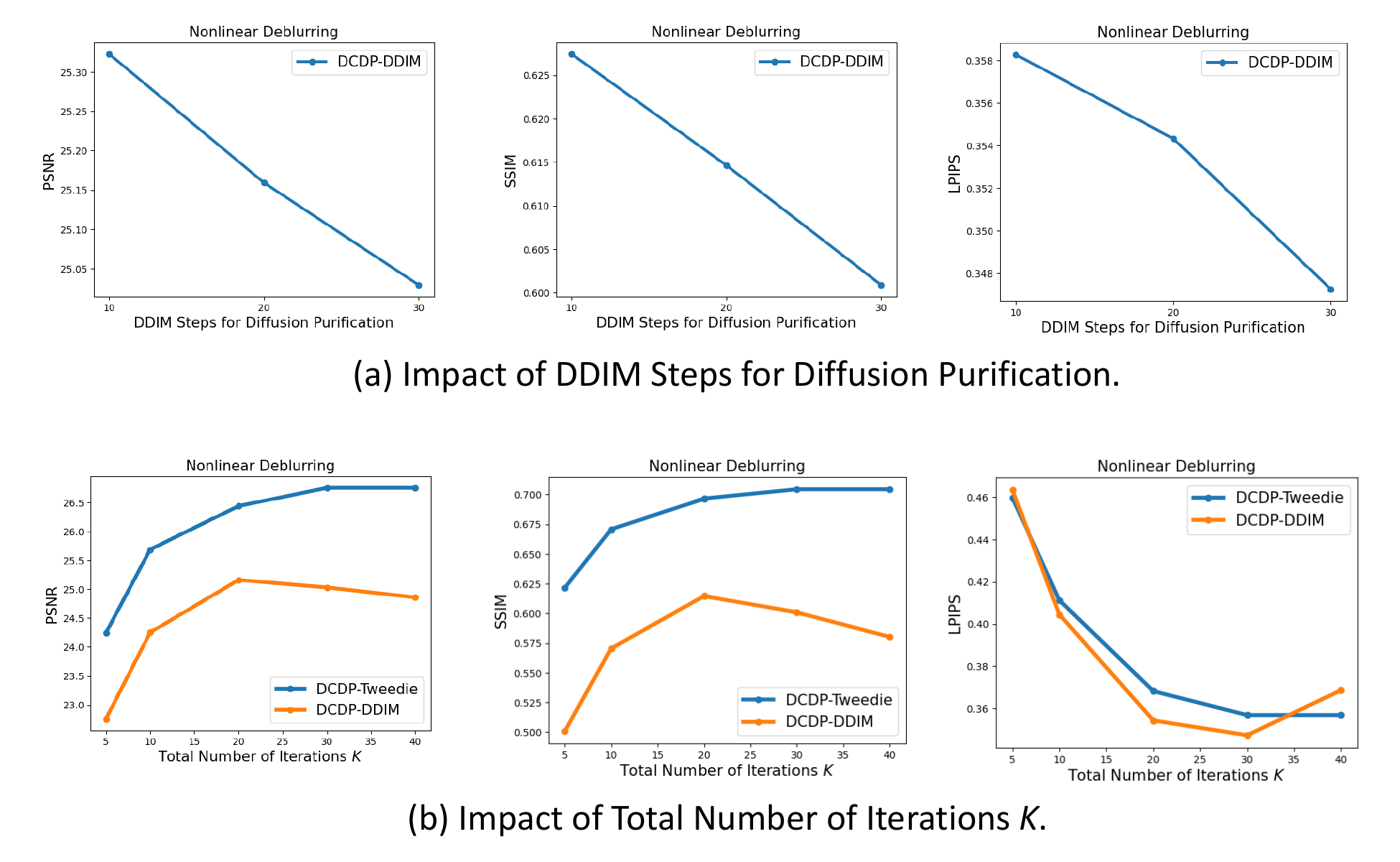}
\caption{\textbf{Impact of number of DDIM steps and total number of iterations $K$.} Figure (a) illustrates the performance of DCDP-DDIM with different numbers of DDIM steps for diffusion purification. Figure (b) illustrates the impact of the total number of iterations $K$ on the performance of DCDP-DDIM and DCDP-Tweedie respectively. }
\label{fig: ablation study on DDIM Steps and total number of iterations.}
\end{figure*}

\medskip
\noindent \textbf{Impact of Number of DDIM Steps.} Next, we investigate how the number of DDIM steps for diffusion purification affects the reconstruction quality. We focus on solving the nonlinear deblurring with pixel space diffusion models (DCDP-DDIM), varying the number of DDIM steps among $\{10, 20, 30\}$, while fixing all other hyperparameters (see~\Cref{app:extra-results} for details). As shown in~\Cref{fig: ablation study on DDIM Steps and total number of iterations.}(a), increasing the number of DDIM steps degrades PSNR and SSIM—metrics that primarily capture data fidelity—while improving the LPIPS score, which better reflects perceptual quality. This reveals a trade-off between fidelity and perceptual quality controlled by the number of DDIM steps. Based on this observation, we fix the number of DDIM steps to 20 in all DCDP-DDIM and DCDP-LDM-DDIM experiments to strike a balance between the two.

\medskip
\noindent \textbf{Impact of Total Number of Iterations $K$.} Lastly, we investigate how the number of total iterations $K$ affects reconstruction quality. Again, we focus on solving the nonlinear deblurring using pixel space diffusion models (both DCDP-DDIM and DCDP-Tweedie), varying $K$ over $\{5,10,20,30,40\}$ while keeping all other hyperparameters fixed (see~\Cref{app:extra-results} for details). In each iteration, we use a fixed number of steps $\tau=100$ for data fidelity optimization, resulting in a total of $\{500, 1000, 2000,3000,4000\}$ data consistency steps. As shown in~\Cref{fig: ablation study on DDIM Steps and total number of iterations.}(b), increasing $K$ initially improves performance across standard metrics, but the gains plateau or even degrade beyond a certain point. In practice, we find that choosing $K$ between 10 and 20 strikes a good balance between reconstruction quality and inference speed.

\begin{figure}[h]
    \centering
    \includegraphics[width=0.9\linewidth]{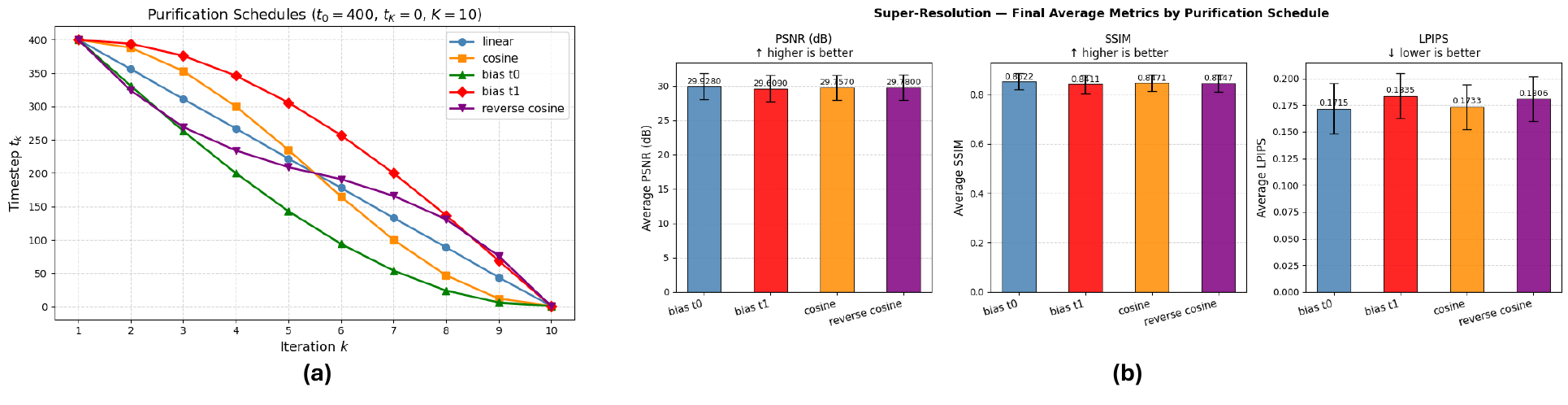}
    \caption{\textbf{Impact of decaying schedules.}
    (a) Different decaying schedules considered in our experiments.
    (b) Super-resolution performance of DCDP under different schedules,
    averaged over 10 randomly selected images.}
    \label{fig:schedule-ablation}
\end{figure}

\medskip
\noindent \textbf{Impact of Decaying Schedules.} As discussed in~\Cref{sec:dcdp}, DCDP achieves robust performance under various decaying schedules. Following~\cite{delbracio2023inversion}, we evaluate several choices on the super-resolution task, including linear, cosine, reverse cosine, biased-$t_0$, and biased-$t_1$ schedules. As shown in~\Cref{fig:schedule-ablation}, these schedules yield comparable performance, suggesting that DCDP is not sensitive to the exact functional form of the schedule.






\section{Conclusion}\label{sec:conclusion}
In this study, we introduce a novel diffusion-based method for solving general inverse problems, which decouples the data consistency steps from the reverse diffusion process.
 Our technique alternates between a reconstruction phase that enforces data consistency and a refinement stage that uses diffusion purification to enforce the diffusion image prior.
 We demonstrated the effectiveness of our algorithm across a wide range of inverse problems, including deblurring, inpainting, and super-resolution. Notably, by decoupling the two processes, we showed that our algorithm can incorporate a broader range of diffusion methods, such as accelerated samplers, latent diffusion models, and consistency models. 

\section*{Acknowledgements}
QQ, XL, SMK acknowledge support from NSF CAREER CCF-2143904, NSF CCF-2212066, NSF CCF-2212326, NSF IIS 2312842, ONR N00014-22-1-2529, an AWS AI Award, MICDE Catalyst Grant, and a gift grant from KLA. SL and SR acknowledge support from NSF CCF-2212065.

\section*{Data Availability Statement}
The code and instructions for reproducing the experiment results will be made available at
\url{https://github.com/Morefre/Decoupled-Data-Consistency-with-Diffusion-Purification-for-Image-Restoration}.

\printbibliography

\appendix
\clearpage


\appendix
\begin{table}[h]

\centering
\begin{tabular}{l|l}
\hline
 \multirow{1}{*}{Notation} & \multicolumn{1}{c}{Definition} \\
\hline
  $K \in \mathbb{N}$ & Number of iterations for overall algorithm \\
\hline
  $\tau \in \mathbb{N}$ & Total number of iterations for data consistency optimization \\
\hline
 $lr \in \mathbb{R}$ & Learning rate for data consistency optimization \\
\hline
 $t_k \in \mathbb{R}$ & Diffusion purification strength at the $k^{\text{th}}
$ iteration \\
\hline
 $\bm v_k \in \mathbb{R}^n$ & Prior consistent sample at $k^{\text{th}}
$ iteration \\
\hline
 $\bm x_k \in \mathbb{R}^n$ & Data consistent sample at $k^{\text{th}}
$ iteration \\
\hline
\end{tabular}
\caption{\textbf{Summary of the notation used throughout this work}}
\label{table:notation}
\end{table} 

\section{Discussion on Hyperparameters}\label{sec: hyperparameters}
Here, we discuss the hyperparameters of our algorithm and some typical choices we make throughout our experiments.

\begin{itemize}[leftmargin=*]
     \item \textbf{Total Number of Iterations.} 
     We run the algorithm for a total of $K$ iterations, which refers to reconstructing the image by alternately applying data fidelity optimization and diffusion purification for $K$ rounds. Empirically, we observe that a value of $K$ ranging from 10 to 20 is sufficient for generating high quality reconstructions. We suggest users start with $K=10$ and then gradually decrease it if a faster computation speed is required or increase it if higher reconstruction quality is required.
     \item \textbf{Optimizer for Data Fidelity Optimization.} Throughout our experiments, we use momentum-based gradient descent with the momentum hyperparameter set to 0.9. The optimal learning rate ($lr$) for data fidelity optimization varies from task to task. In general, $lr$ should neither be too large, which will not result in images with meaningful structures, nor too small, in which case the convergence speed of the overall algorithm will be too slow. Throughout the experiments we fix the number of gradient steps for each round of the data fidelity optimization. When the measurements are corrupted by noise, one can further perform early stopping to prevent overfitting to the noise.
     \item \textbf{Decaying Time Schedule for Diffusion Purification.} The decaying time schedule plays the key role in the success of our proposed algorithm. As described in ~\Cref{sec:dcdp}, given a data-fidelity-based reconstruction $\mb x_k$, the goal of diffusion purification is to generate $\mb v_k$ that shares the overall structural and semantic information of $\mb x_k$, such that it is consistent with the measurements while at the same time being free from the artifacts visible in $\mb x_k$. Unfortunately, the $\mb v_k$ resulting from diffusion purification is sensitive to the choice of $t_k$, i.e., the amount of noise added to $\mb x_k$. When $t_k$ is small, the resulting $\mb v_k$ will exhibit similar structure to $\mb x_k$ but with artifacts, as there is not enough noise to submerge these artifacts through diffusion purification. On the other hand, when $t_k$ is too large,  though $\mb v_k$ will be free from artifacts, its semantic information and overall structure will significantly deviate from $\mb x_k$. As a result, the resulting reconstructions will not be measurement-consistent.
     
     To this end, previous works that utilize diffusion purification~\cite{alkhouri2023robust, choi2021ilvr} depend on fine-tuning an optimal $t^\star$. However, this fine-tuning strategy can be infeasible since the optimal $t^\star$ varies from task to task and even from image to image. To alleviate the need to identify the optimal $t^\star$, we implement a linearly decaying schedule $\{t_k\}_{k\in [K]}$ for each iteration $k$. In the early iterations when $t_k$ is relatively large, the resulting $\mb v_k$, though inconsistent with the measurements, will be free from artifacts and can serve as a good starting point for the subsequent iteration of data fidelity optimization. As $t_k$ keeps decreasing, the algorithm will converge to an equilibrium point which is both consistent with the measurements and free from the artifacts. When the measurements are noiseless, we observe that for a diffusion model trained with 1000 total timesteps, a linearly decaying time schedule starting from $t_1=400$ and ending at $t_{K}=0$ results in high-quality reconstructions for Super-Resolution, Gaussian Deblurring, and Motion Deblurring. For Box Inpainting, we choose a higher initial timestep $t_1=700$ and $t_1=500$ for pixel space diffusion models and LDMs, respectively, to compensate for the severe artifacts in the large box region. When the measurements are corrupted by noise, one may instead end at $t_{K}=\tilde{t}$, where $\tilde{t}>0$, to compensate for the negative effect of the measurement noise. We demonstrate the reconstruction process for various tasks in~\Cref{fig: SR_Process,fig: GD_Process,fig: Inpainting_Process,fig: MD_Process,fig:Measurements,fig:ProcessClean,fig:ProcessNoisy}. Notice that the diffusion purification removes the artifacts from the initial reconstructions obtained via data fidelity optimization, especially in the early iterations, and results in high-quality reconstructions consistent with the measurements as the timestep decreases.
    \begin{figure*}[t!]
    \centering    
    \includegraphics[width=\textwidth]{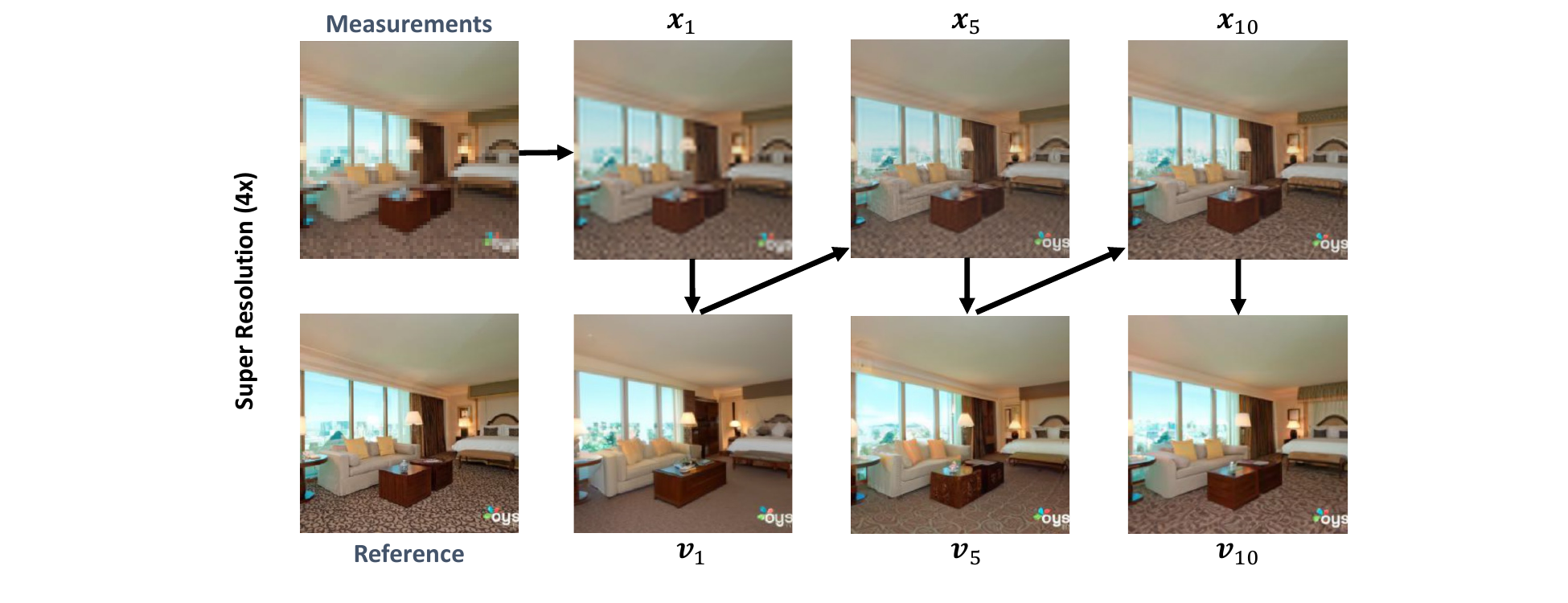}
    \caption{\textbf{Reconstruction process for super-resolution with pixel space diffusion models.} The reconstruction process takes $K=10$ iterations in total with a time schedule starting at $t_1=400$ and linearly decaying to $t_{10}=0$. We demonstrate reconstructions at timesteps 1, 5, and 10.  }
    \label{fig: SR_Process}
    \end{figure*}
    
    \begin{figure*}[t!]
    \centering    
    \includegraphics[width=\textwidth]{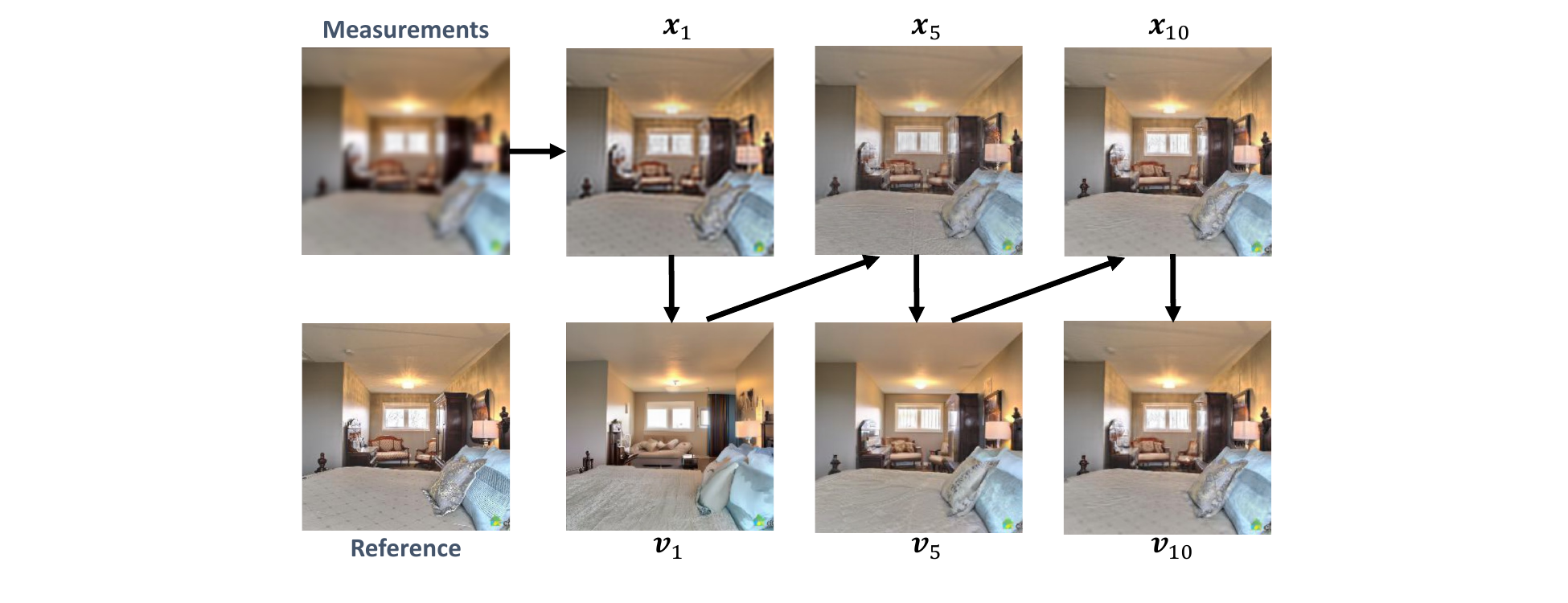}
    \caption{\textbf{Reconstruction Process for Gaussian Deblurring with Pixel Space Diffusion Models.} The reconstruction process takes $K=10$ iterations in total with a time schedule starting at $t_1=400$ and linearly decaying to $t_{10}=0$. Due to the space limit, only reconstructions at timesteps 1, 5 and 10 are shown. }
    \label{fig: GD_Process}
    \end{figure*}

    \begin{figure*}[t!]
    \centering    
    \includegraphics[width=\textwidth]{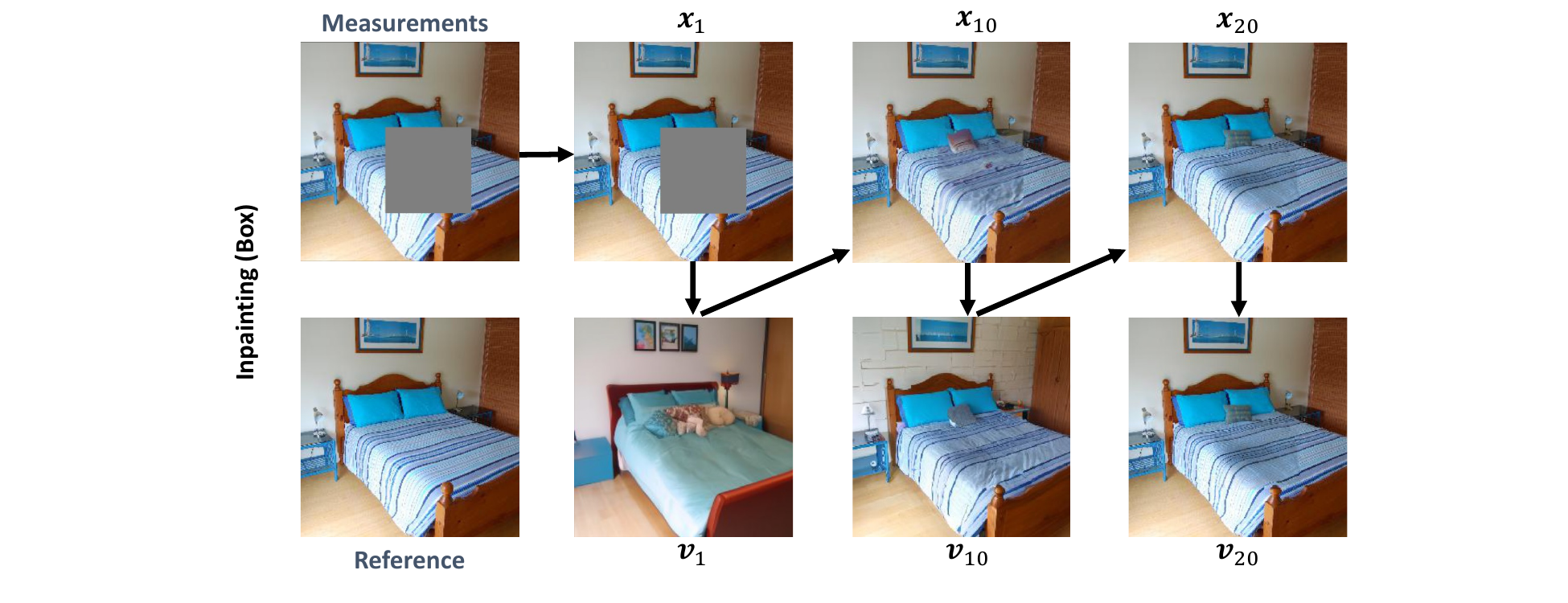}
    \caption{\textbf{Reconstruction Process for Inpainting with Pixel Space Diffusion Models.} The reconstruction process takes $K=20$ iterations in total with a time schedule starting at $t_1=700$ and linearly decaying to $t_{20}=0$. Due to the space limit, only reconstructions at timesteps 1, 10 and 20 are shown.}
    \label{fig: Inpainting_Process}
    \end{figure*}
    \begin{figure*}[t!]
    \centering    
    \includegraphics[width=\textwidth]{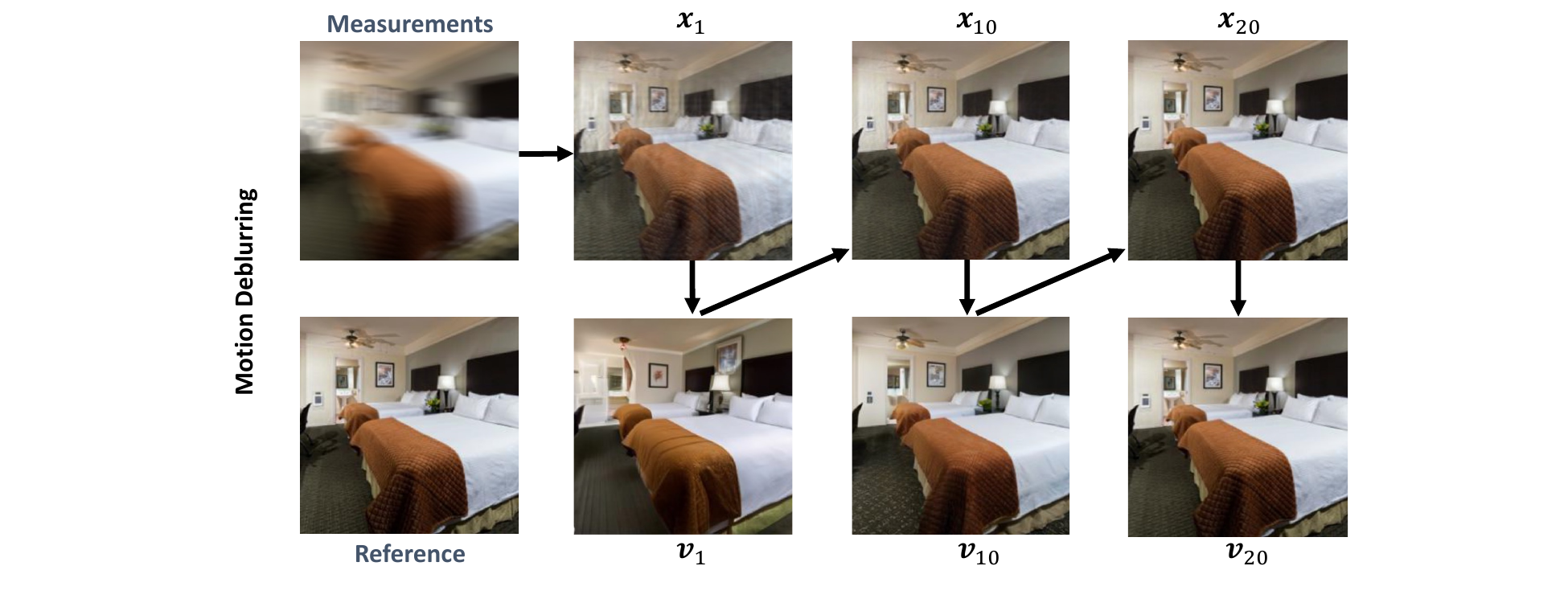}
    \caption{\textbf{Reconstruction Process for Motion Deblurring with Pixel Space Diffusion Models.} The reconstruction process takes $K=20$ iterations in total with a time schedule starting at $t_1=400$ and linearly decaying to $t_{20}=0$. Due to the space limit, only reconstructions at timesteps 1, 10 and 20 are shown.}
    \label{fig: MD_Process}
    \end{figure*}
    
     \item \textbf{DDIM Steps for Diffusion Purification.}
     With the help of accelerated diffusion samplers such as DDIMs \cite{song2020denoising}, we can perform diffusion purification with as few as 20 unconditional sampling steps. As mentioned before, the total number of iterations $K$ ranging from 10-20 is sufficient for generating high quality reconstructions. This amounts to 200 to 400 unconditional reverse sampling steps in total, a significantly lower figure compared to DPS~\cite{chung2022diffusion}, which requires 1000 unconditional reverse sampling steps. Interestingly, we find that increasing the DDIM sampling steps does not necessarily improve the performance of diffusion purification. Therefore, we suggest users use 20 steps for diffusion purification.
     \item \textbf{Pixel Space Optimization and Latent Space Optimization.} As discussed in~\Cref{sec: adaptation to LDMs}, in the case of LDMs, one can use either Pixel Space Optimization~\eqref{eqn:data-ldm-pixel}  or Latent Space Optimization~\eqref{eqn:data-ldm} to enforce data consistency. Depending on the specific task, one approach may generate $\mb x_k$ with higher quality than the other. In practice, as shown in \Cref{fig: Latent_VS_Pixel_MD,fig: Latent_VS_Pixel_GD}, for motion deblurring and Gaussian deblurring, the initial reconstructions generated with Pixel Space Optimization have significantly higher quality than those generated with Latent Space Optimization. Furthermore, Pixel Space Optimization is much faster than Latent Space Optimization since the gradient computation of the former approach does not require backpropagation through the deep decoder $\mathcal{D}(\cdot)$. To achieve the best reconstruction quality, we suggest users try both options and choose the one which gives the best performance. On the other hand, for applications that require fast reconstruction speed, we suggest using Pixel Space Optimization.
\end{itemize}

    \begin{figure*}[t!]
    \centering    
    \includegraphics[width=\textwidth]{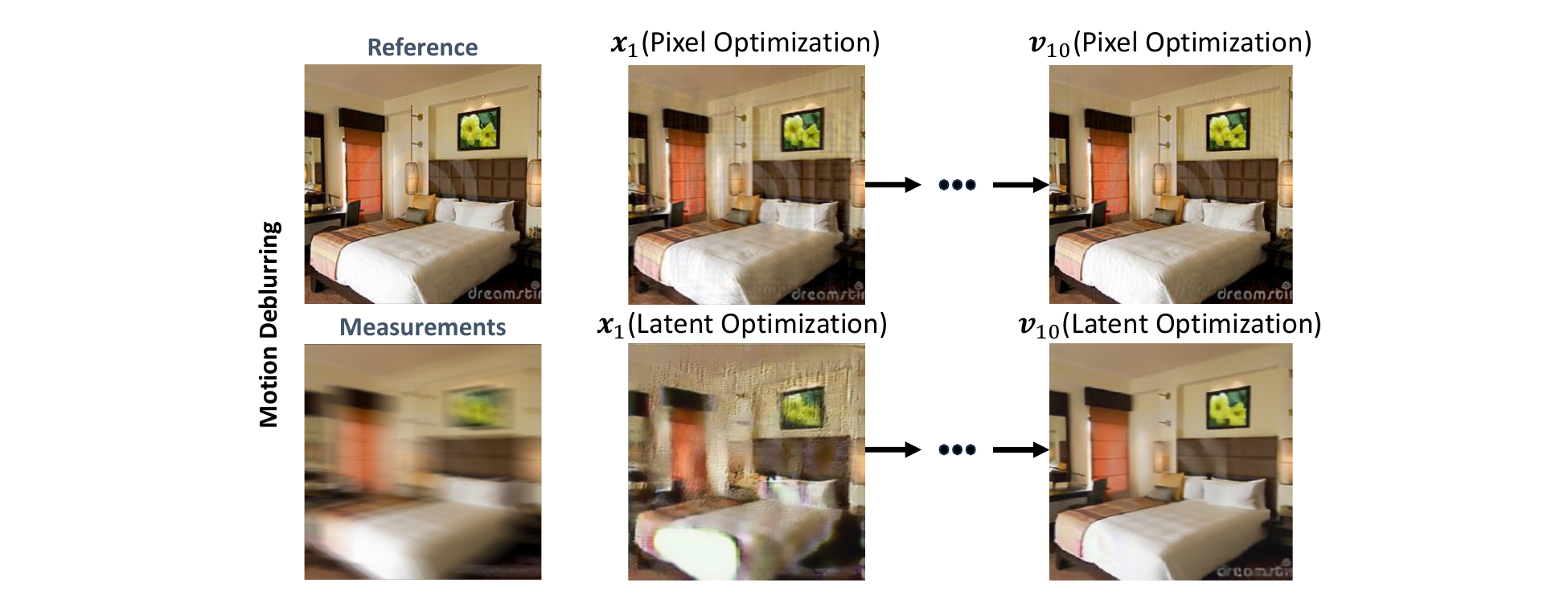}
    \caption{\textbf{Comparison between Pixel Space Optimization and Latent Space Optimization for Motion Deblurring.} Initial reconstructions obtained via Pixel Space Optimization have much better quality compared to the ones obtained from Latent Space Optimization which in turn result in final reconstructions with higher quality.}
    \label{fig: Latent_VS_Pixel_MD}
    \end{figure*}
    
    \begin{figure*}[t!]
    \centering    
    \includegraphics[width=\textwidth]{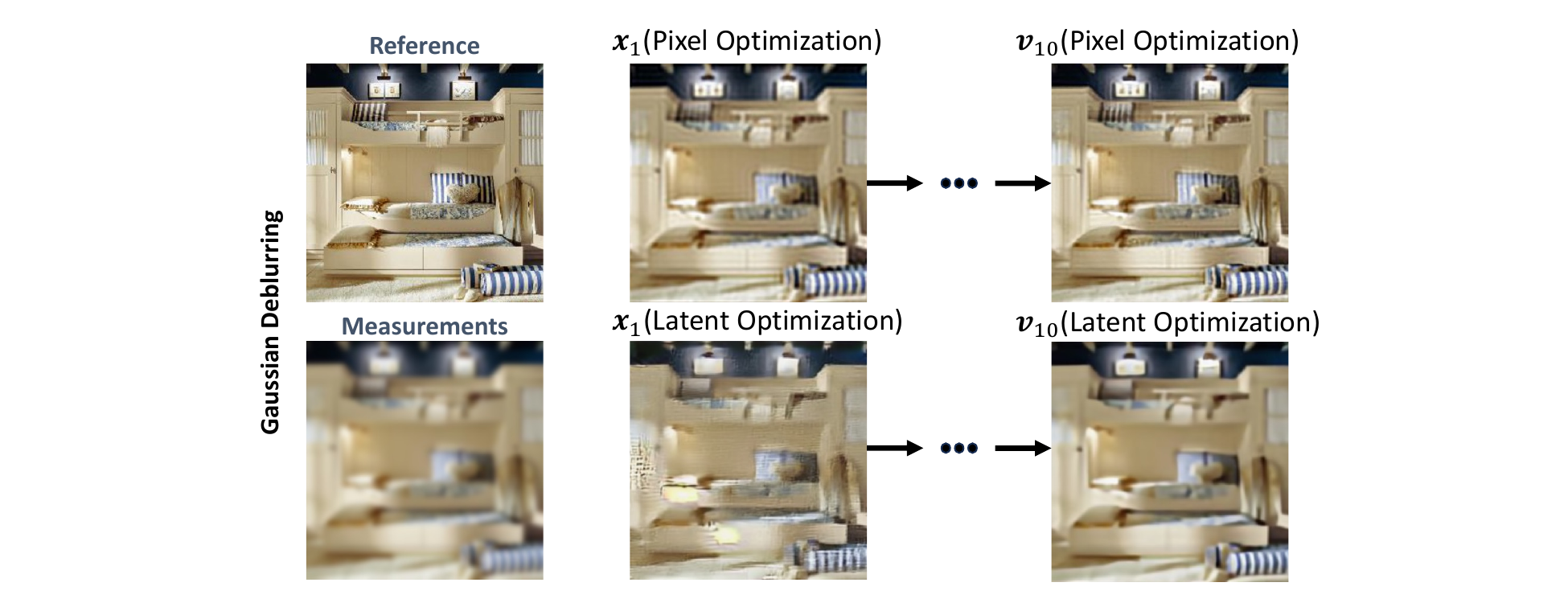}
    \caption{\textbf{Comparison between Pixel Space Optimization and Latent Space Optimization for Gaussian Deblurring.} Initial reconstructions obtained via Pixel Space Optimization have much better quality compared to the ones obtained from Latent Space Optimization which in turn result in final reconstructions with higher quality.}
    \label{fig: Latent_VS_Pixel_GD}
    \end{figure*}

\section{Choice of Hyperparameters}\label{choice of hyperparameters}

In this section, we specify the hyperparameters of DCDP for generating the results in Section~\ref{sec: experiments}. For the baselines, we tune the hyperparameters via grid search to obtain the best possible performance.

\paragraph{Experiments with Pixel Space Diffusion Models}
\label{sec: pixel space hyper}

For the pixel space experiments, we tabulate the hyperparameters for each dataset in~\Cref{table: pixel_hyperparameters} (linear inverse problems) and in~\Cref{table: nonlinear deblurring hyperparameters} (nonlinear deblurring). Recall that for $t_k$, we use a linearly decaying time schedule, which starts at an initial value $t_1$ and ends at $t_K$. For linear inverse problems, the DCDP-DDIM and DCDP-Tweedie share the same set of hyperparameters. The diffusion purification in DCDP-DDIM is performed with 20 deterministic DDIM steps. 

We optimize the data fidelity objective with momentum-based gradient descent with the momentum hyperparameter set to $0.9$ for all tasks. For linear inverse problems, we fix the total number of gradient steps as 1000 throughout the $K$ iterations, which means $\tau=\frac{1000}{K}$ in each round of data fidelity optimization. For nonlinear deblurring, we instead use a total number of 2000 gradient steps, i.e., $\tau=\frac{2000}{K}$ in each round of data fidelity optimization. 

\setlength{\tabcolsep}{2.5pt}
\begin{table*}[t!]

\centering
\begin{tabular}{l|ccccc|ccccc|ccccc|ccccc}
\hline
 \multirow{2}{*}{Dataset} & \multicolumn{5}{c|}{Super Resolution ($4\times$)} & \multicolumn{5}{c|}{Gaussian Deblurring} & \multicolumn{5}{c|}{Motion Deblurring} & \multicolumn{5}{c}{Inpainting (Box)} \\

 & $lr$ & $K$ &$\tau$ & $t_1$ &$t_K$  & $lr$ & $K$ &$\tau$ & $t_1$ &$t_K$  & $lr$ & $K$ &$\tau$ & $t_1$ &$t_K$ & $lr$ & $K$ &$\tau$ & $t_1$ &$t_K$\\ 
\hline
LSUN-Bedroom  & $10^{3}$& $10$ & $100$ & $400$& $0$
&$10^{5}$ & $10$& $100$ &$400$& $0$ &$10^{5}$& $20$ &$50$& $400$ & $0$ &$10^{3}$ &$20$& $50$ & $700$ &$0$ \\
FFHQ  & $10^{3}$& $10$ & $100$ & $400$& $0$ & $10^{5}$ & $10$&$100$ & $400$&$0$ & $10^{5}$& $20$ &$50$ & $400$ & $0$& $10^{3}$ & $20$ & $50$ & $700$ &$0$ \\
\hline
\end{tabular}
\caption{\textbf{Hyperparameters of DCDP with pixel space diffusion models for solving linear inverse problems.} For linear inverse problems, DCDP-Tweedie and DCDP-DDIM share the same set of hyperparameters.}
\label{table: pixel_hyperparameters}
\end{table*}

\setlength{\tabcolsep}{2.5pt}
\begin{table*}[t!]

\centering
\begin{tabular}{l|ccccc|ccccc|ccccc|ccccc}
\hline
 \multirow{2}{*}{Dataset} & \multicolumn{5}{c|}{Super Resolution ($4\times$)} & \multicolumn{5}{c|}{Gaussian Deblurring} & \multicolumn{5}{c|}{Motion Deblurring} & \multicolumn{5}{c}{Inpainting (Box)} \\

  & $lr$ & $K$ &$\tau$ & $t_1$ &$t_K$  & $lr$ & $K$ &$\tau$ & $t_1$ &$t_K$  & $lr$ & $K$ &$\tau$ & $t_1$ &$t_K$ & $lr$ & $K$ &$\tau$ & $t_1$ &$t_K$\\ 

\hline
LSUN-Bedroom  & $10^{3}$& $10$& $100$ & $400$& $0$ &$10^{5}$ & $10$& $100$ & $400$& $0$ &$10^{5}$ & $10$ & $100$ & $400$ & $0$&$10^{3}$ & $20$ & $50$ & $500$ & $0$ \\
FFHQ  & $10^{3}$& $10$& $100$ & $400$& $0$ & $10^{5}$ & $4$& $250$ & $400$& $0$ & $10^{5}$& $10$ & $100$ & $400$ & $0$ & $10^{3}$ & $20$ & $50$ & $500$ & $0$ \\
\hline
\end{tabular}
\caption{\textbf{Hyperparameters of DCDP with latent space diffusion models for solving linear inverse problems.} For linear inverse problems, DCDP-Tweedie and DCDP-DDIM share the same set of hyperparameters.}
\label{table: latent_hyperparameters}
\end{table*}

\setlength{\tabcolsep}{2.5pt}
\begin{table*}[t!]

\centering
\begin{tabular}{l|ccccc|ccccc|ccccc|ccccc}
\hline
 \multirow{2}{*}{Dataset} & \multicolumn{5}{c|}{DCDP-DDIM} & \multicolumn{5}{c|}{DCDP-Tweedie} & \multicolumn{5}{c|}{DCDP-LDM-DDIM} & \multicolumn{5}{c}{DCDP-LDM-Tweedie} \\

  & $lr$ & $K$ &$\tau$ & $t_1$ &$t_K$  & $lr$ & $K$ &$\tau$ & $t_1$ &$t_K$  & $lr$ & $K$ &$\tau$ & $t_1$ &$t_K$ & $lr$ & $K$ &$\tau$ & $t_1$ &$t_K$\\ 

\hline
ImageNet  & $10^{3}$& $20$& $100$ & $400$& $100$ &$10^{3}$ & $20$& $100$ & $400$& $50$ &$5\times10^{3}$ & $20$ & $100$ & $300$ & $50$&$5\times10^{3}$ & $20$ & $100$ & $300$ & $50$ \\
\hline
\end{tabular}
\caption{\textbf{Hyperparameters of DCDP for solving nonlinear deblurring.}}
\label{table: nonlinear deblurring hyperparameters}
\end{table*}

\setlength{\tabcolsep}{2.5pt}
\begin{table*}[t!]
\centering
\begin{tabular}{l|ccccc|ccccc|ccccc|ccccc}
\hline
 \multirow{2}{*}{Dataset} & \multicolumn{5}{c|}{Super Resolution ($4\times$)} & \multicolumn{5}{c|}{Gaussian Deblurring} & \multicolumn{5}{c|}{Motion Deblurring} & \multicolumn{5}{c}{Inpainting (Box)} \\

& $lr$ & $K$ &$\tau$ & $t_1$ &$t_K$  & $lr$ & $K$ &$\tau$ & $t_1$ &$t_K$  & $lr$ & $K$ &$\tau$ & $t_1$ &$t_K$ & $lr$ & $K$ &$\tau$ & $t_1$ &$t_K$\\
\hline
LSUN-Bedroom  & $10^{3}$& $20$&$50$ & $1$& $0$ & $10^{5}$ & $20$& $50$ & $1$& $0$& $10^{5}$& $20$ & $50$ & $1$ & $0$ & $10^{3}$ & $20$ & $50$ & $5$ & $0$\\
\hline
\end{tabular}
\caption{\textbf{Hyperparameters of DCDP with CMs.} Different from the diffusion models in previous experiments which specify the noise level via discrete forward steps (ranging from 0 to 1,000), CMs are trained on a continuous range of Gaussian noise with standard deviation spanning [0.002,80]. Therefore, $t_1$ and $t_K$ in this table stand for the standard deviation of the noise rather than the timestep.}
\label{table: CMs_hyperparameters}
\end{table*}

\setlength{\tabcolsep}{5pt}
\begin{table*}[t!]
\begin{center}
\resizebox{0.95\textwidth}{!}{%
\begin{tabular}{l|lll|lll}
\hline
 \multirow{2}{*}{Method} & \multicolumn{3}{c|}{Super Resolution $4\times$} & \multicolumn{3}{c}{Inpainting (Box)} \\

& LPIPS$\downarrow$ & PSNR$\uparrow$ & SSIM$\uparrow$  & LPIPS$\downarrow$ & PSNR$\uparrow$ & SSIM$\uparrow$  \\

\hline
DPS~\cite{chung2022diffusion} & $\underline{0.176} $ {$ \pm 0.04$} & $28.45 \pm 1.99$ & $0.792 \pm 0.05$ & $0.115 \pm 0.02$ & $23.14\pm 3.14$ & $0.829 \pm 0.03$ \\
MCG~\cite{chung2022improving}  & $0.182 \pm 0.03$ & $24.45 \pm 2.93$ & $0.736 \pm 0.06$ &$0.119 \pm 0.02$ & $22.77 \pm 3.06$ & $0.827 \pm 0.03$\\
ADMM-PnP~\cite{ahmad2020plug} & $0.307 \pm 0.03$ & $21.54 \pm 3.12$ & $0.766 \pm 0.03$   & $0.251 \pm 0.02$ & $18.26 \pm 1.86$ & $0.724 \pm 0.02$ \\
DDNM~\cite{wang2022zero} & $0.197 \pm 0.04$ & $28.03 \pm 1.46$ & $0.795 \pm 0.08$   & $0.235 \pm 0.06$ & $24.47 \pm 2.38$ & $0.837 \pm 0.08$\\

RED-Diff~\cite{mardani2024a} & $0.212 \pm 0.06$ & $27.18 \pm 1.23$ & $0.781 \pm 0.07$  & $0.165 \pm 0.08$& $24.22 \pm 3.16$ & $0.824 \pm 0.05$\\

DiffPIR~\cite{zhu2023denoising} & $0.260 \pm 0.06$ & $26.64 \pm 1.78$ & $ 0.767 \pm 0.06$   & $0.108 \pm 0.07$& $23.89 \pm 1.89$ & $0.814 \pm 0.06$ \\
DCDP-DDIM &$\bm{0.154} \pm 0.03$ & $\underline{29.43} \pm 2.01$& $\underline{0.844} \pm 0.03$ & $\underline{0.052} \pm 0.01$ & $\underline{25.54} \pm 3.01$  & $\underline{0.917} \pm 0.01$  \\

DCDP-Tweedie &$0.185 \pm 0.03$ & $\bm{30.86} \pm 2.11 $& $\bm{0.878} \pm 0.03$ & $\bm{0.051} \pm 0.01$ & $\bm{28.02} \pm 2.74$  & $\bm{0.935} \pm 0.01$  \\

\hline
 \rowcolor{Columnbia Blue}
  PSLD~\cite{rout2024solving}  & $0.309 \pm 0.06$  & $26.22 \pm 2.07$  & $0.758 \pm 0.06$   &$0.248 \pm 0.06$ & $25.42 \pm 2.15$ & $0.788 \pm 0.05 $ \\

\rowcolor{Columnbia Blue}
  DCDP-LDM-DDIM &$\bm{0.183} \pm 0.04$ & $\underline{29.16} \pm 1.65$& $\underline{0.826} \pm 0.03$ & $\bm{0.166} \pm 0.03$ & $\underline{26.50} \pm 2.20$  & $\underline{0.860} \pm 0.02$  \\
\rowcolor{Columnbia Blue}
  DCDP-LDM-Tweedie &$\underline{0.215} \pm 0.04$ & $\bm{30.61} \pm 1.97$& $\bm{0.866} \pm 0.03$ & $\underline{0.204} \pm 0.04$ & $\bm{27.47} \pm 2.05$ & $\bm{0.868} \pm 0.03$ \\
\hline
\end{tabular}}
\end{center}
\caption{\textbf{Quantitative results of super resolution and inpainting on the FFHQ dataset for both pixel-based diffusion models and LDMs.} The results colored in \textcolor{Columnbia Blue}{blue} are for LDMs, while the others are for pixel space diffusion models. The best results are in bold and second best results are underlined for pixel space models and LDMs  respectively.}
\label{table: quant3}
\end{table*}

\setlength{\tabcolsep}{5pt}
\begin{table*}[t!]
\begin{center}
\resizebox{\textwidth}{!}{%
\begin{tabular}{l|lll|lll}
\hline
 \multirow{2}{*}{Method} & \multicolumn{3}{c|}{Gaussian Deblurring} & \multicolumn{3}{c}{Motion Deblurring}  \\

& LPIPS$\downarrow$ & PSNR$\uparrow$ & SSIM$\uparrow$  & LPIPS$\downarrow$ & PSNR$\uparrow$ & SSIM$\uparrow$ \\

\hline
 DPS~\cite{chung2022diffusion} & $\underline{0.162} \pm 0.03$ & $27.97 \pm 2.01$ & $0.782 \pm 0.05$ & $0.219 \pm 0.04$ &  $25.34 \pm 2.21$ & $0.690 \pm 0.07$  \\

  MCG~\cite{chung2022improving} & $0.164 \pm 0.03$ & $25.03 \pm 3.62$ & $0.790 \pm 0.06$ & $0.343 \pm 0.04$ &  $24.88 \pm 2.01$ & $0.708 \pm 0.05$  \\
DDNM~\cite{wang2022zero} & $0.217 \pm 0.04$ & $28.23 \pm 1.69$ & $0.805 \pm 0.08$   &-- & -- & --\\

RED-Diff~\cite{mardani2024a} & $0.242 \pm 0.06$ & $25.25 \pm 2.17$ & $0.791 \pm 0.05$  & $0.268 \pm 0.08$& $24.32 \pm 2.67$ & $0.754 \pm 0.05$\\

DiffPIR~\cite{zhu2023denoising} & $0.236 \pm 0.06$ & $27.34 \pm 2.15$ & $ 0.778 \pm 0.04$   & $0.255 \pm 0.07$& $26.69 \pm 1.49$ & $0.856 \pm 0.06$ \\
  ADMM-PnP~\cite{ahmad2020plug} &$0.421 \pm 0.03$ & $20.86 \pm 2.16$& $0.664 \pm 0.05$ & $0.522 \pm 0.02$ & $17.91 \pm 1.99$  & $0.547 \pm 0.06$  \\

  DCDP-DDIM &$\bm{0.140} \pm 0.02$ & $\underline{30.41} \pm 1.93$& $\underline{0.864} \pm 0.03$ & $\bm{0.051} \pm 0.01$ & $\underline{37.56} \pm 1.77$  & $\underline{0.963} \pm 0.01$  
  \\
  
  DCDP-Tweedie &$0.173 \pm 0.03$ & $\bm{31.67} \pm 2.12$& $\bm{0.891} \pm 0.03$ & $\underline{0.060} \pm 0.01$ & $\bm{38.37} \pm 1.88$  & $\bm{0.970} \pm 0.01$  \\


\hline
  \rowcolor{Columnbia Blue}
  PSLD~\cite{rout2024solving}  & $0.270 \pm 0.05$ & $27.46 \pm 1.84$  & $0.790 \pm 0.05$  & $0.379 \pm 0.05$ & $22.35 \pm 1.85$ & $0.641 \pm 0.07$ \\
  \rowcolor{Columnbia Blue}
  DCDP-LDM-DDIM 
  &$\underline{0.198} \pm 0.03$ & $\underline{28.91} \pm 1.98$ & $\underline{0.825} \pm 0.04$ & $\underline{0.152} \pm 0.03$ & $\underline{30.11} \pm 1.93$  & $\underline{0.851} \pm 0.03$ \\ 
\rowcolor{Columnbia Blue}
  DCDP-LDM-Tweedie &$\bm{0.171} \pm 0.02$ & $\bm{30.45} \pm 1.86$& $\bm{0.877} \pm 0.03$ & $\bm{0.108} \pm 0.02$ & $\bm{31.48} \pm 1.94$  & $\bm{0.894} \pm 0.02$  \\
\hline
\end{tabular}}
\end{center}
\caption{\textbf{Quantitative results of Gaussian and motion deblurring on the FFHQ dataset for both pixel-based diffusion models and LDMs.} The results colored in \textcolor{Columnbia Blue}{blue} are for LDMs, while the others are for pixel space diffusion models. The best results are in bold and second best results are underlined for pixel space models and LDMs respectively.}
\label{table: quant4}
\end{table*}

\paragraph{Experiments with LDMs}
We specify the hyperparameters used for experiments using LDMs in Table~\ref{table: latent_hyperparameters} (linear inverse problems) and in~\Cref{table: nonlinear deblurring hyperparameters} (nonlinear deblurring). Again, DCDP-LDM-DDIM and DCDP-LDM-Tweedie share the same set of hyperparameters for linear inverse problems. The diffusion purification in DCDP-LDM-DDIM is performed with 20 deterministic DDIM steps and for the data fidelity objectives, the same procedure used in the pixel space is applied to the LDMs.
To achieve the best performance, we use Pixel Space Optimization for Gaussian deblurring and motion deblurring while using Latent Space Optimization for box inpainting, super-resolution and nonlinear deblurring. Again, we fix the total number of gradient steps for linear inverse problems as 1000 while using 2000 steps for nonlinear deblurring.

\paragraph{Experiments with CMs}
Lastly, we specify the hyperparameters used for experiments using CMs in Table~\ref{table: CMs_hyperparameters}. Note that different from the typical diffusion models which specify the noise level via discrete forward steps (commonly ranging from 0 to 1,000), CMs are trained on a continuous range of Gaussian noise with standard deviation spanning [0.002,80].

\section{Dealing with Measurement Noise}\label{sec: robust study detail}
\begin{figure*}[t]
    \centering
    \begin{minipage}{\textwidth}
        \centering
        \includegraphics[width=0.7\textwidth]{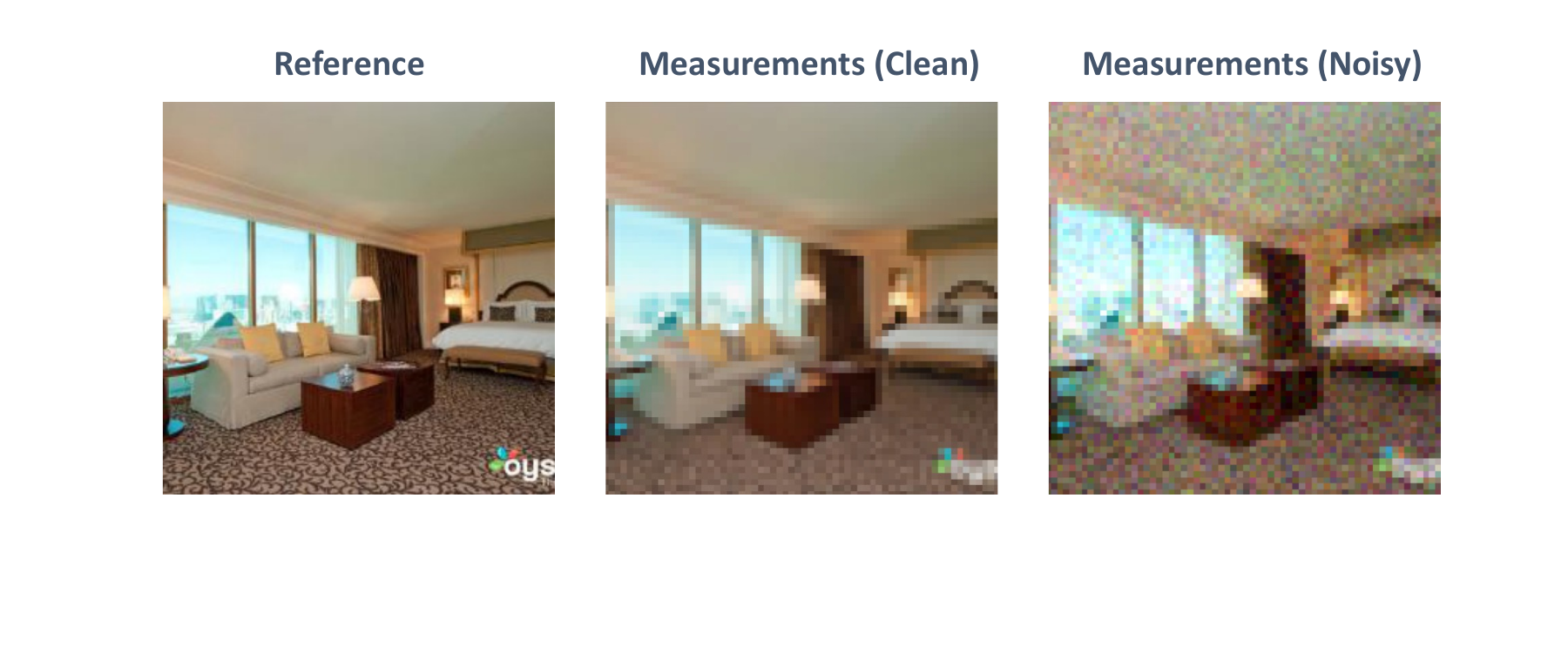}
        \vspace{-0.4in}
        \caption{\textbf{Reference and Measurements for Super-Resolution (4x).} The noisy measurements are corrupted by random Gaussian noise with $\sigma=0.1$.}
        \label{fig:Measurements}
    \end{minipage}

    \begin{minipage}{\textwidth}
        \centering
        \includegraphics[width=1\textwidth]{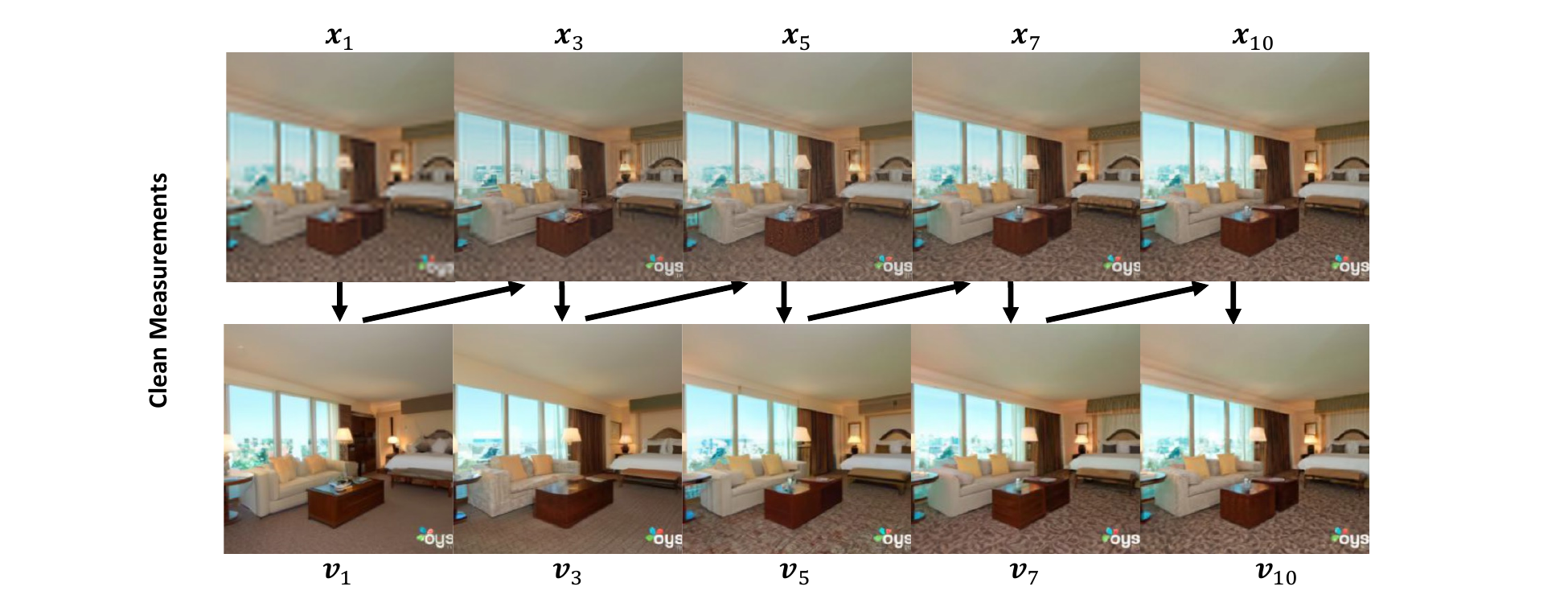}
        \caption{\textbf{Reconstruction Process with Clean Measurements.} The reconstruction process takes $K=10$ iterations in total with a time schedule starting at $t_1=400$ and linearly decaying to $t_{10}=0$. The learning rate for data fidelity optimization is $lr=1000$. Due to the space limit, only reconstructions at timesteps 1, 3, 5, 7 and 10 are shown.}
        \label{fig:ProcessClean}
    \end{minipage}

    \begin{minipage}{\textwidth}
        \centering
        \includegraphics[width=1\textwidth]{figures/Progress_2.pdf}
        \caption{\textbf{Reconstruction Process with Noisy Measurements.} To compensate the artifacts introduced by the measurement noise, the reconstruction process takes $K=10$ iterations in total with a time schedule starting at $t_1=400$ and linearly decaying to $t_{10}=200$. The learning rate for data fidelity optimization is $lr=100$.}
        \label{fig:ProcessNoisy}
    \end{minipage}
    \label{fig:whole_process}
\end{figure*}

When the measurements are noiseless, a linearly decaying schedule starting from $t_1$ and ending at $t_{K}=0$ is capable of resulting in high-quality reconstructions, and we can use a relatively high learning rate for the data fidelity optimization. On the contrary, in the presence of measurement noise, the data fidelity optimization in~\eqref{Pixel Space Data Fidelity Simple} will inevitably introduce additional artifacts even if $\mb v_{k-1}$ is the ground truth image. To achieve robust recovery, one may instead: \emph{(i)} use a relatively small learning rate for data fidelity optimization to prevent overfitting to the noise, and \emph{(ii)} end the time schedule at $t_{K}=\tilde{t}$, where $\tilde{t}>0$, to compensate for the artifacts resulting from the measurement noise. As shown in~\Cref{fig:ProcessClean,fig:ProcessNoisy}, compared to the clean measurements case, due to the additive noise, the data fidelity optimization always introduces artifacts in $\mb x_k$. This issue can be alleviated by utilizing a non-zero diffusion purification strength at the final iteration. For the experiments in~\Cref{subsec:exp-ablation}, all the hyperparameters are kept the same as described in \Cref{choice of hyperparameters}, but with the following changes:
\begin{itemize}
    \item The learning rate for motion deblurring changes from $1e^5$ to $1e^4$.
    \item The time schedules for super-resolution and motion deblurring start from $t_1=400$ and decay linearly to $250$ and $300$ respectively.
\end{itemize}
With the changes made above, our algorithm achieves reconstruction performance comparable to that of DPS with measurement noise. Notice that the hyperparameters are chosen such that they work under the most severe noise level, i.e., $\sigma=0.1$. Better performance can be achieved by decreasing $t_K$ in the case of $\sigma=0.05$ or $\sigma=0.075$.

\section{Additional Experiment Results and Implementation Details}\label{app:extra-results}
\medskip
\noindent \textbf{Quantitative Results on FFHQ.} 
Quantitative results for solving inverse problems on the FFHQ dataset are presented in~\Cref{table: quant3,table: quant4}. Note that DCDP outperforms the baselines for various inverse problems. 

\begin{figure*}[t]
    \centering
    \includegraphics[width=0.9\linewidth]{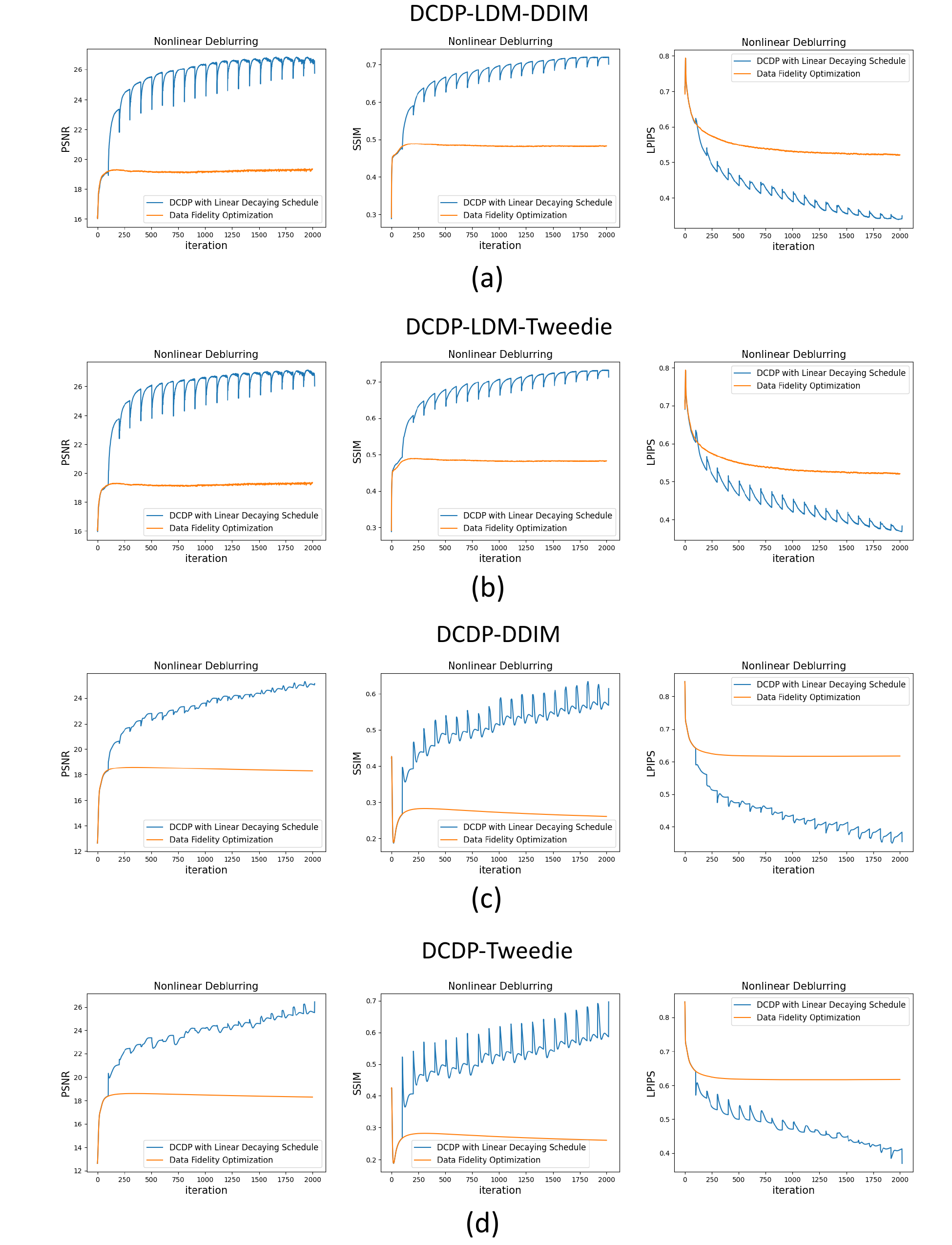}
    \caption{\textbf{Convergence of our methods in terms of PSNR, SSIM and LPIPS for solving nonlinear deblurring.} DCDP greatly improves upon the pure data fidelity optimization by leveraging the image priors embedded in the diffusion models.}
    \label{fig:dcdp convergence behavior nonlinear deblur.}
\end{figure*}

\medskip
\noindent \textbf{Convergence Behavior.} In~\Cref{subsec:exp-ablation} (\Cref{fig: Convergence}), we show that with a linearly decaying time schedule for diffusion purification, DCDP iteratively improves the LPIPS scores and progressively converges towards a high-quality reconstruction. Similar convergence behavior holds when measuring the reconstruction quality using PSNR and SSIM, as shown in~\Cref{fig:dcdp convergence behavior nonlinear deblur.}. In this experiment, DCDP is applied to the nonlinear deblurring task, with the standard metrics averaged over 10 randomly selected images from the ImageNet dataset. The hyperparameters are the same as those reported in~\Cref{choice of hyperparameters} (\Cref{table: nonlinear deblurring hyperparameters}). Note that DCDP substantially improves upon the pure data fidelity optimization, implying diffusion purification is an efficient operator to enforce the image priors embedded in the pre-trained diffusion models.

\medskip
\noindent\textbf{Impact of Number of DDIM Steps.} Here we provide experimental details for the ablation study on the impact of DDIM steps in~\Cref{subsec:exp-ablation}. The experiments are conducted using a pixel space diffusion model based on DCDP (DCDP-DDIM) for solving the nonlinear deblurring problem. Except for the number of DDIM steps, all other hyperparameters are the same as those reported in~\Cref{choice of hyperparameters} (\Cref{table: nonlinear deblurring hyperparameters}). The standard metrics are averaged over 10 randomly selected images from the ImageNet dataset.

\medskip
\noindent \textbf{Impact of Total Number of Iterations $K$.} 
Here we provide experimental details for the ablation study on the impact of $K$ in~\Cref{subsec:exp-ablation}. The experiments are conducted using pixel space diffusion models (both DCDP-DDIM and DCDP-Tweedie). Except for $K$, all other hyperparameters are the same as those reported in~\Cref{choice of hyperparameters} (\Cref{table: nonlinear deblurring hyperparameters}). The standard metrics are averaged over 10 randomly selected images from the ImageNet dataset.




\end{document}